\documentclass[preprint,12pt]{elsarticle}

\usepackage{amssymb,color}

\journal{Journal of Computational Physics}

\begin{document}

\begin{frontmatter}



\title{Improvements to the Levenberg-Marquardt algorithm for nonlinear least-squares minimization}


\author[lassp]{Mark K. Transtrum}
\author[lassp]{James P. Sethna}

\address[lassp]{Laboratory of Atomic and Solid State Physics, Cornell University, Ithaca, New York 14853, USA}

\begin{abstract}
When minimizing a nonlinear least-squares function, the Levenberg-Marquardt algorithm can suffer from a slow convergence, particularly when it must navigate a narrow canyon en route to a best fit.  On the other hand, when the least-squares function is very flat, the algorithm may easily become lost in parameter space.  We introduce several improvements to the Levenberg-Marquardt algorithm in order to improve both its convergence speed and robustness to initial parameter guesses.  We update the usual step to include a geodesic acceleration correction term, explore a systematic way of accepting uphill steps that may increase the residual sum of squares due to Umrigar and Nightingale, and employ the Broyden method to update the Jacobian matrix.  We test these changes by comparing their performance on a number of test problems with standard implementations of the algorithm.  We suggest that these two particular challenges, slow convergence and robustness to initial guesses, are complimentary problems.  Schemes that improve convergence speed often make the algorithm less robust to the initial guess, and vice versa.  We provide an open source implementation of our improvements that allow the user to adjust the algorithm parameters to suit particular needs.
\end{abstract}


\end{frontmatter}



\section{Introduction}
\label{LM:sec:introduction}

A common computational problem is that of minimizing a sum of squares
\begin{equation}
  \label{LM:eq:Cost}
  C(\theta) = \frac{1}{2}\sum_{m=1}^M r_m(\theta)^2,
\end{equation}
where $ r:\mathbf{R}^N \rightarrow \mathbf{R}^M$ is an $M$-dimensional nonlinear vector function of $N$ parameters, $\theta$, where $M \geq N$.  The Levenberg-Marquardt algorithm is perhaps the most common method for nonlinear least-squares minimization.  In this paper, we discuss a number of modifications to the Levenberg-Marquardt algorithm designed to improve both its success rate and convergence speed.  These modifications are likely to be most useful on large problems with many parameters, where the usual Levenberg-Marquardt routine often has difficulty.

Least-squares minimization is most often used in data fitting, in which case the function $r_m(\theta)$ takes the form
\begin{equation}
  \label{LM:eq:residual}
  r_m(\theta) = \frac{y(t_m,\theta)-y_m}{\sigma_m},
\end{equation}
where $y(t,\theta)$ is a model of the observed data, $y_m$, that depends on a set of unknown parameters, $\theta$, and one or more independent variables $t$.  The deviation of the model from observation is weighted by the uncertainty in observed data, $\sigma$.  The terms in Eq.~(\ref{LM:eq:residual}) are known as the residuals and may be augmented by additional terms representing Bayesian prior information about the expected values of $\theta$.  We refer to the function in Eq.~(\ref{LM:eq:Cost}) as the cost function.  The cost corresponds to the negative $\log$-likelihood of parameter values given the data assuming Gaussian errors.  The parameter values that minimize $C(\theta)$ are known as the best fit parameters.

\begin{figure}[htbp]
  \centering
  \includegraphics[width=6in]{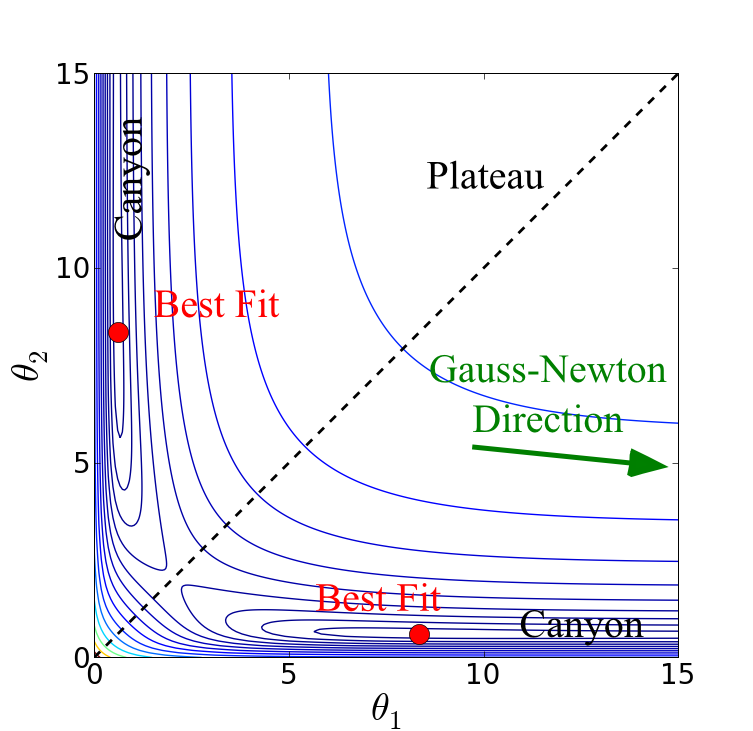}
  \caption[Cost contours in parameter space]{\label{LM:fig:Contours} The cost surface in parameter space for least-squares problems often forms a hierarchy of narrow, winding canyons surrounded by plateaus.  Algorithms are easily lost on the plateaus, often evaporating parameters (pushing them to infinity) while searching for a canyon.  We see here that the Gauss-Newton direction in the plateau region is nearly orthogonal to the ideal direction.  Having found the canyon, algorithms can become sluggish while following it to the best fit.  This simple model, $y=e^{-\theta_1t} + e^{-\theta_2t}$, fit to three data points has a plateau when the parameters become very large\cite{Transtrum2010}.  It also exhibits a symmetry when parameter are permuted.}
\end{figure}

The Levenberg-Marquardt algorithm\cite{Levenberg1944, Marquardt1963, More1977,Press2007} is is a modification of the Gauss-Newton method, which is is based on a local linearization of the residuals
\begin{equation}
  \label{LM:eq:reslinear}
  r_m(\theta + \delta \theta) \approx r_m(\theta) + J_{m\mu}\delta \theta^\mu,
\end{equation}
where $J$ is the Jacobian matrix $J_{m\mu}=\partial r_m / \partial \theta_\mu$.  The Gauss-Newton method then iterates according to
\begin{equation}
  \label{LM:eq:GNstep}
  \delta \theta = -(J^TJ)^{-1} \nabla C = - (J^TJ)^{-1} J^T r.
\end{equation}
The Gauss-Newton method will usually converge quickly if it begins sufficiently near a minimum of $C$.  However, the matrix $J^TJ$ is often ill-conditioned, with eigenvalues often spanning a range of six orders of magnitude or more.  Therefore, unless the initial guess is very good, the Gauss-Newton method takes large, uncontrolled steps and will fail to converge.  This is illustrated explicitly in figure \ref{LM:fig:Contours}, where far from the best fit the Gauss-Newton direction is nearly orthogonal to the direction the algorithm ought to take.

To remedy the shortcomings of the Gauss-Newton method, Levenberg and Marquardt each suggested damping the $J^TJ$ matrix by a diagonal cutoff\cite{Levenberg1944,Marquardt1963}.  The Levenberg-Marquardt algorithm therefore steps according to
\begin{equation}
  \label{LM:eq:LMstep}
  \delta \theta = -\left( J^TJ + \lambda D^TD \right)^{-1} \nabla C.
\end{equation}
where $D^TD$ is a positive-definite, diagonal matrix representing the relative scaling of the parameters and $\lambda$ is a damping parameter adjusted by the algorithm.  When $\lambda$ is large, the method takes a small step in the gradient direction.  As the method nears a solution, $\lambda$ is chosen to be small and the method converges quickly via the Gauss-Newton method.

Often, models with many parameters exhibit universal characteristics known as sloppiness which pose particular challenges to the fitting process.  The behavior of sloppy models is determined by only a few stiff (relevant) parameter combinations, while most other parameter combinations are sloppy (irrelevant)\cite{Brown2003, Brown2004, Mortensen2005, Waterfall2006, Gutenkunst2007a}.  Fitting difficulties arise when algorithms are lost in regions of parameter space where the model behavior is insensitive to changes in the parameters, i.e.~a plateau on the cost surface in parameter space as in Fig.~\ref{LM:fig:Contours}.  A common occurrence is that while lost on the plateau, algorithms push parameters to infinite values without finding a good fit, a phenomenon known as parameter evaporation\cite{Transtrum2010,Transtrum2011}.  Although these solutions correspond to fixed points of the cost, i.e.~$\nabla C = 0$, and are therefore either local minima or saddle points at infinity, these solution are unsatisfactory since they often correspond to bad fits to the data and the parameters, being infinite, have little meaning.  The algorithm must then be guided by hand in order to find a better fit.  This problem can sometimes be avoided by augmenting the cost function with penalty terms to keep the parameter within a reasonable range as suggested in reference\cite{Transtrum2011}.  However, this is not always possible, and penalty terms will move the location of the minimum.  One would therefore like the algorithm itself to be less sensitive to parameter evaporation.

In addition to parameter evaporation, the algorithm becomes sluggish when it must follow a narrow canyon to find the best fit, as in Fig.~\ref{LM:fig:Contours}.  It is common for the aspect ratio of the canyon to be greater than $1000:1$ for problems with ten or more parameters\cite{Waterfall2006, Gutenkunst2007a}, which requires the algorithm to take very small steps as it inches along the bottom of the trough.  The difficulty in data fitting is exacerbated by the fact that solutions to the two principal problems (parameter evaporation and slow convergence) are often at odds with one another.  Methods which speed up convergence in the canyon usually increase the probability of parameter evaporation and vice versa.

Because it is can tune $\lambda$ as needed, the Levenberg-Marquardt method is well-suited for dealing with the difficulties in nonlinear least-squares minimization.  By properly adjusting the damping term, the method can interpolate between gradient descent, for avoiding parameter evaporation, and the Gauss-Newton algorithm for quickly converging along a canyon. One of the challenges for the Levenberg-Marquardt method is in choosing a suitable scheme for updating the damping parameter $\lambda$ that successfully interpolates between the two regimes.  Many such schemes exist and although some are more suited for avoiding parameter evaporation and others are more adept at navigating the canyon, the Levenberg-Marquardt method is generally robust to the specific method used.

Its relative success notwithstanding, the Levenberg-Marquardt algorithm may still fail to converge if it begins far from a minimum and will converge slowly if it must inch along the bottom of a canyon.  Given the ubiquitous role of nonlinear least-squares minimization in mathematical modeling, and considering the trend to use increasingly large and computationally expensive models in all areas of science and engineering, any improvements that could be made to the Levenberg-Marquardt algorithm would be welcome.  In this paper we discuss several such improvements.

This paper is organized as follows: In section \ref{LM:sec:damping} we summarize the key elements of the Levenberg-Marquardt algorithm.  We then explore how the existing methods can be improved by including the so-called geodesic acceleration\cite{Transtrum2010,Transtrum2011} in section \ref{LM:sec:geodesic} and a modified acceptance criterion due to Umrigar and Nightingale\cite{UmrigarBold} in section \ref{LM:sec:uphill}.  We then discuss how a rank-deficient update to the Jacobian matrix can reduce the number of times it must be evaluated in section \ref{LM:sec:Broyden}.  An open source implementation of the Levenberg-Marquardt algorithm with our proposed improvements is available in FORTRAN\cite{SFGeoLM}.

In each of the following sections we compare the performance of the algorithm with the suggested improvements on a set of several test problems drawn from the Minpack-2 project\cite{Averick1992} and the NIST statistical reference datasets\cite{mccullough1998}.  Because most of these problems are of small or moderate size, most are much easier than the larger, more difficult fitting problems that motivated our work.  In order to make these problems more difficult, we test the algorithm for an ensemble of starting points drawn from a broad distribution.  We find that for a sufficiently diverse set of starting points, these problems can be made of comparable difficulty to larger, more challenging problems. We also explore the algorithms' performance on several large test problems drawn from recent research.  Because the Minpack-2 and NIST problems can be evaluated quickly, they make ideal test problems provided the more difficult starting positions are used.  These problems are summarized in \ref{LM:sec:testprobs}.  We find that our proposed improvements consistently improve the performance of the algorithm on these problems.

\section{The Levenberg-Marquardt algorithm}
\label{LM:sec:damping}

In this section we describe the basic concepts of the Levenberg-Marquardt algorithm.  Our implementation of the Levenberg-Marquardt algorithm consists of iteratively repeating the following five steps:
\begin{enumerate}
\item  Update the function and Jacobian values (if necessary) based on the current parameter values.
\item  Update the scaling matrix $D^TD$ and damping parameter $\lambda$.
\item  Propose a parameter step, $\delta \theta$, and evaluate the function at the new parameter values, $\theta + \delta \theta$.
\item  Accept or reject the parameter step depending on whether the cost has decreased at the new parameters.
\item  Stop if the algorithm has met any of the desired convergence criteria or has exceeded the limit of function evaluations or iterations.
\end{enumerate}
The only ambiguities in this method are in the method for selecting $\lambda$ and the scaling matrix $D^TD$.  We discuss specific methods for selecting $\lambda$ in section \ref{LM:sec:lambda} and in section \ref{LM:sec:DTD} we discuss how to select $D^TD$.   Finally we discuss convergence and stopping criteria in section \ref{LM:sec:convergence}.

In later sections we discuss how to modify the other aspects of the Levenberg-Marquardt algorithm.  These improvements represent our contribution to the algorithm, and we find that these improvements can offer drastic improvements to its speed and stability.  In particular, as mentioned above, we modify the proposed parameter step, $\delta \theta$, in section \ref{LM:sec:geodesic} to include a second order correction that we call the geodesic acceleration; in section \ref{LM:sec:uphill} we modify how the algorithm accepts the proposed step; and in section \ref{LM:sec:Broyden} we modify how the Jacobian matrix is updated.  

\subsection{Choosing the damping parameter}
\label{LM:sec:lambda}

The basic strategy behind choosing the damping term uses the observations that the square of the step size $\Delta^2 = \delta \theta^T D^TD \delta \theta$ is a monotonically decreasing function of $\lambda$ in Eq.~(\ref{LM:eq:LMstep}).  Therefore, for a sufficiently large value of $\lambda$, the algorithm will take an arbitrarily small step in a descent direction.  If a proposed step is unacceptable, one need only increase the damping term until a smaller, more acceptable step has been found.  Because choosing $\lambda$ is equivalent to choosing the step size, the Levenberg-Marquardt method can be considered a trust-region method.  There are two broad classes of methods for determining the appropriate damping.  This can be done by either adjusting $\lambda$ directly, or, by first choosing an acceptable step size $\Delta$ and then finding a $\lambda$ such that $|\delta \theta| \leq \Delta$ (note that reference \cite{More1977} describes how $\lambda$ may be efficiently found for a given $\Delta$).  We will refer to these two types of schemes as direct and indirect methods respectively.

Many schemes have been developed to efficiently adjust $\lambda$ or $\Delta$.  In our experience, the simple method originally suggested by Marquardt (with a slight modification described shortly) is usually adequate.  In this scheme, if a step is accepted, then $\lambda$ is decreased by a fixed factor, say $10$.  If a step is rejected then $\lambda$ is appropriately raised by a factor of $10$.  As explained in reference \cite{Transtrum2011}, the qualitative effect of the damping term is to modify the eigenvalues of the matrix $J^TJ + \lambda D^TD$ to be at least $\lambda$.   Often, the eigenvalues of $J^TJ$ are well spaced on a $\log$-scale; it is therefore natural to choose the factor by which $\lambda$ is either raised/lowered to be comparable to the eigenvalue spacing of $J^TJ$.  We have much greater success using a factor of $2$ or $3$ on most problems.  Additionally, we find that lowering $\lambda$ by a larger factor than it is raised also produces more beneficial results.  For many moderate sized problems decreasing by a factor of $3$ and raising by a factor of $2$ is adequate.  For larger problems, decreasing by a factor of $5$ and raising by a factor of $1.5$ is better.  This scheme is known as delayed gratification, and its motivation is described in \cite[section VIII]{Transtrum2011}.  This basic strategy can also be applied to an indirect method by systematically increasing/decreasing $\Delta$ by a multiplicative factor instead of $\lambda$.  

We compare the relative performance of these methods for selecting $\lambda$, together with more complicated schemes describe by Nielson \cite{nielsen1999} and Mor\'{e} \cite{More1977} on the problems in \ref{LM:sec:testprobs}.  In our experiments, no single method consistently outperforms all other methods on all the test problems, with the results depending very strongly on which problem one considers.  We present our results of these tests in \ref{LM:sec:standardcomparison}.  

\begin{figure}[htbp]
  \centering
  \includegraphics[width=6in]{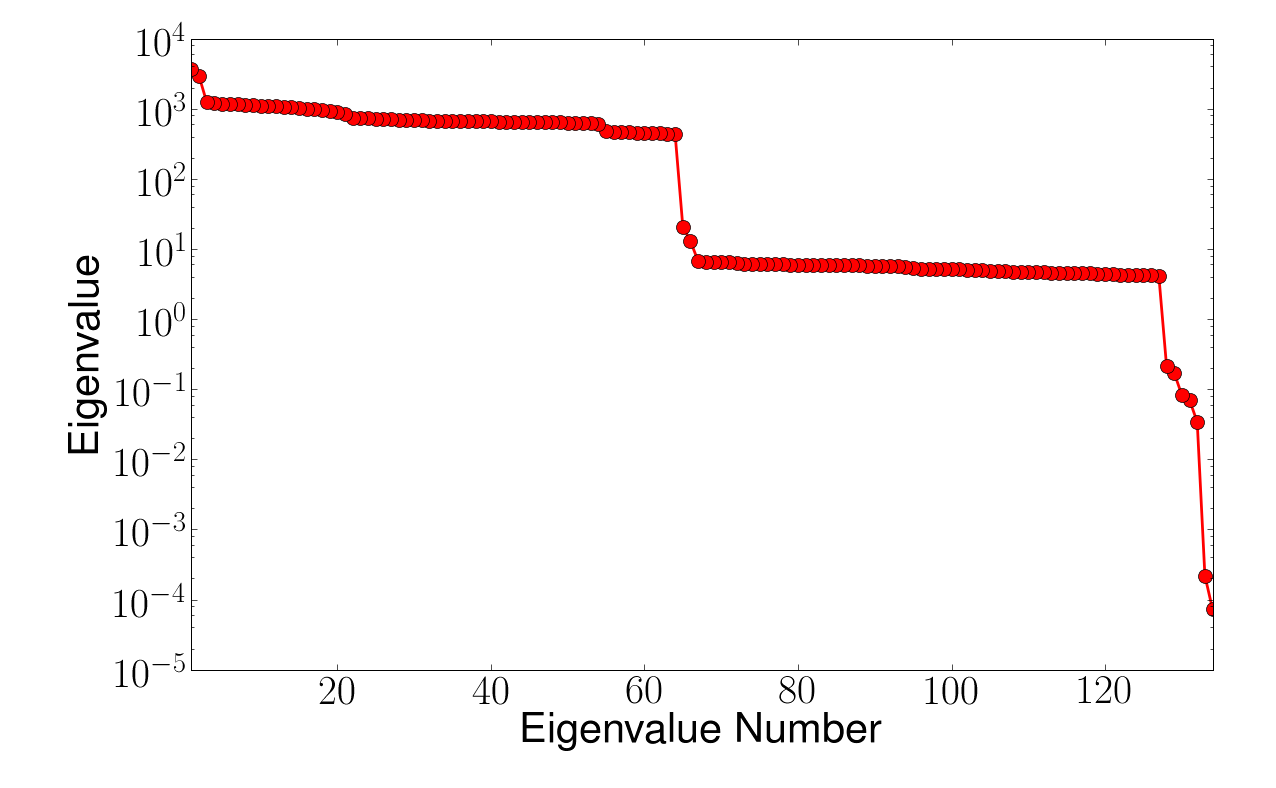}
  \caption[Eigenvalues for problem C]{The eigenvalues of of the Hessian matrix for problem C.  Although the eigenvalues span nearly eight order of magnitude, they are not evenly spaced over this range.  There are two large collections are eigenvalues near $1000$ and between $1$ and $10$.  This clumping helps explain why indirect update methods are superior for this problem.}
  \label{LM:fig:ctseigs}
\end{figure}

By inspecting the results in  \ref{LM:sec:standardcomparison} more closely, we find there are problems for which indirect methods collectively outperform the direct methods and vice versa.  We can understand these trends by considering the problems individually.  For example, the relative success of indirect methods on problems B and C can be understood by considering the eigenvalues of the of the Hessian matrix $J^TJ$ near the best fit, as in figure \ref{LM:fig:ctseigs}.  Notice that although the eigenvalues span many order of magnitude, they tend to collect near $10^3$ and $10^1$.  As we argued above, $\lambda$ should be turned based on the spacing of these eigenvalues.  In order to be effective, an algorithm would therefore need to carefully tune $\lambda$ while it was near $10^3$, but then change very quickly while between $10^1$ and $10^3$.  The direct methods described above cannot do this, however, the indirect methods that tune $\Delta$ rather than $\lambda$ seem to accomplish this naturally.

The relative success of direct methods on other problems can be understood by a similar argument.  In these problems there are regions of parameter space in which the step size must be finely tuned and other regions in which must change by large amounts.  It seems for these problems the direct methods that tune $\lambda$ rather than the step size are more efficient.


\subsection{Damping Matrix}
\label{LM:sec:DTD}

We now describe how to choose an effective damping matrix $D^TD$.  Levenberg originally suggested an additive damping strategy, corresponding to $D^TD = \delta$, the identity.  It has since been suggested that a multiplicative damping strategy in which $D^TD$ is a diagonal matrix with entries equal to the diagonal entries of $J^TJ$ would more accurately capture the scaling of the several parameters.  This method for choosing parameter scaling has the property that the algorithm is invariant to rescaling parameters, which is to say if the parameters of the model were replaced by the parameters $\tilde{\theta}_i = \lambda_i \theta_i$ for some scaling factors $\lambda_i$, then the sequence of iterates, i.e. values of the cost function, produced by the algorithm would be left unchanged.

The problem with a method that preserves scale invariance is that it greatly increases the susceptibility to parameter evaporation.  In particular, if a parameter begins to evaporate, the model becomes less sensitive to the parameter, so it's corresponding entry on the diagonal of $J^TJ$ becomes small, in turn decreasing the damping of this parameter.  This, however, is exactly the wrong behavior for dealing with parameter evaporation.  Indeed, as argued previously\cite{Transtrum2011} the role of the damping matrix is to introduce parameter dependence to the step, so a choice that is scale invariant is somewhat counter productive.  On the other hand, we find that using the Marquardt scaling can greatly speed up the algorithm when it is in the region of a canyon, when scale independence is crucial.

The popular implementation of Levenberg-Marquardt found in Minpack uses a similar but superior method described in reference \cite{More1977}.  It chooses $D^TD$ to be diagonal with entries given by the largest diagonal entries of $J^TJ$ yet encountered.  This method also preserves invariance under rescaling but is more robust at avoiding parameter evaporation, however, it is still more prone to parameter evaporation than Levenberg scaling.  This is because initial parameter guesses may lie in regions that do not produce enough damping.

A good compromise is to specify a minimum value of the damping terms in $D^TD$.  This prevents the damping from being too small, either if they begin far from the canyon or as they parameters evaporate, but allows the algorithm to fine tune the scaling as it follows the canyon.  We find that this method is both robust to parameter evaporation and efficient at finding good fits.

\subsection{Convergence Criteria}
\label{LM:sec:convergence}

Finally, we discuss criteria for the algorithm to stop searching for a best fit.  It is important to distinguish between convergence criteria and stopping criteria.  The former are criteria indicating that the algorithm has indeed found a local minimum of the cost, while the latter are criteria indicating that the algorithm is, in effect, giving up.  In our comparison of the several algorithms, we consider the convergence rate as one measure of the algorithm's performance (see \ref{LM:sec:successmeasures}).  This rate is the fraction of times the algorithm claimed to have successfully found a minimum.

An elegant convergence criterion proposed by Bates and Watts originated in the geometric interpretation of the least-squares problem\cite{Bates1981b}.  This method monitors the angle between the residual vector and the tangent plane, which we denote by $\phi$.  In particular, the cosine of the angle in data space is given by
\begin{equation}
  \label{LM:eq:cosalpha}
  \cos \phi = \frac{|P^T r|}{|r|},
\end{equation}
where $P^T$ is a projection operator that projects into the tangent plane of the model manifold.  Given a singular value decomposition of the Jacobian matrix $J = U \Sigma V^T$, then $P^T = U U^T$.  The algorithm can then be stopped when $\cos \phi$ is less than some quantity, say $10^{-2}$ or $10^{-3}$.

This method provides a dimensionless convergence criterion that indicates how near one is to the minimum.  It also has a statistical interpretation in terms of the accuracy of the solution in terms of the statistical uncertainty in the parameters.  This method has a serious deficiency, however, when the model manifold has narrow boundaries as described in references\cite{Transtrum2010, Transtrum2011}.  If the best fit happens to have evaporated parameters, a likely scenario for large models fit to noisy data, then $\cos \phi$ may be large although the algorithm has in fact converged.

As parameters evaporate, the model becomes less sensitive to that particular parameter combination and the Jacobian matrix has a singular value that becomes vanishingly small.  (Note that the singular values correspond to the square root of the eigenvalues of  $J^T J$.) When it is sufficiently small we should consider these parameter directions to lie in the null space of $J$.  Although the singular value may be formally nonzero, for computational purposes we understand that the algorithm will not make any more progress by moving the parameters in these directions and they should not contribute to the tangential component of the residuals.

To remedy this situation, we replace the projection operator $P^T = U U^T$ in Eq.~(\ref{LM:eq:cosalpha}) with $P^T = \tilde{U} \tilde{U}^T$ where $\tilde{U}$ is a matrix of left singular vectors of $J$ for which the corresponding singular value is larger than some threshold.  If the function is evaluated to precision $\epsilon$, then we find that ignoring the directions with singular values less than $\sqrt{\epsilon} \max{\Sigma}$, where $\max{\Sigma}$ is the largest singular value, works well.  An alternative solution is to use a convergence criteria when the gradient of the cost falls below a certain threshold.

In addition to the convergence criterion described above, the algorithm should have a number of stopping criteria.  In our implementation we provide stopping criteria for when a maximum number of residual and Jacobian evaluations have been reached, in addition to a maximum number of iterations of the algorithm.  We also provide stopping criteria for when the gradient of the cost has fallen to some threshold, when the change in parameter values becomes sufficiently small, the damping term becomes too large, and when cost itself has reached some acceptable value.

\section{Geodesic Acceleration}
\label{LM:sec:geodesic}

In order to improve the efficiency of the Levenberg-Marquardt method, we propose modifying the step to include higher order corrections.  To derive this correction, consider the minimization problem of finding the best residuals with a constrained step-size.  We write the dependence of the residual on the shift $\delta \theta$ as
\begin{equation}
r(\theta + \delta\theta) =  r + J \delta\theta 
	+ 1/2\, \delta\theta^T K \delta\theta + \cdots,
\end{equation}
where $J$ and $K$ are the arrays of first and second derivatives respectively.  We wish to minimize
\begin{equation}
  \min_{\delta\theta} \ \ \left( r + J \delta\theta 
	+ 1/2\, \delta \theta^T K \delta \theta \right)^2
\end{equation}
with the constraint that $\delta \theta^T D^T D \delta \theta \leq \Delta^2$.  After introducing a Lagrange multiplier $\lambda$ for the constraint in the step size, the minimization becomes
\begin{equation}
  \min_{\delta \theta}  \left( r + J \delta \theta
	+ 1/2\, \delta \theta^T K \delta \theta \right)^2 
	+ \lambda \delta \theta^T D^T D \delta \theta.
\end{equation}

By varying $\delta \theta$ we find the normal equations:
\begin{equation}
  \label{LM:eq:normal}
  J_{m\mu} r_m + \left( J_{m\mu} J_{m\nu} + r_m K_{m\mu\nu} + \lambda D_{m\mu} D_{m\nu} \right) \delta \theta^\nu + \left( J_{m\nu} K_{m\mu\alpha} + 1/2\, J_{m\mu} K_{m\nu\alpha} \right) \delta \theta^\nu \delta \theta^\alpha = 0,
\end{equation}
where we have explicitly included all the indices to avoid any ambiguity and used the convention that all repeated indices are summed.  Since we constrain the step size, it is natural to assume that $\delta \theta$ is small, and we seek a solution of Eq.~(\ref{LM:eq:normal}) as a perturbation series around the linearized equation:
\begin{equation}
  \delta \theta = \delta \theta_1 + \delta \theta_2 + \cdots.
\end{equation}
Let $\delta \theta_1$ be a solution of the linearized equation:
\begin{eqnarray}
  \delta \theta_1 & = & -(J^TJ + r^T K + \lambda D^TD)^{-1} J^T r \nonumber\\
  & \approx & -(J^TJ + \lambda D^TD)^{-1} J^T r, \nonumber
\end{eqnarray}
where in the second line we have made the usual Gauss-Newton approximation.  It will turn out that this neglected term will help to cancel out a higher order correction.  We therefore assume
\begin{equation}
  \label{LM:eq:dtheta1}
  \delta \theta_1 = -(J^TJ + \lambda D^TD)^{-1} J^T r.
\end{equation}
is the usual Levenberg-Marquardt step.

With this definition of $\delta \theta_1$, Eq.~(\ref{LM:eq:normal}) becomes
\begin{equation}
  \label{LM:eq:normal2}
   \left( J_{m\mu} J_{m\nu} + r_m K_{m\mu\nu} + \lambda D_{m\mu} D_{m\nu} \right) \delta \theta_2^\nu + 1/2\, J_{m\mu} K_{m\nu\alpha} \delta \theta_1^\nu \delta \theta_1^\alpha + \left( r_m K_{m\mu\alpha}  + \delta \theta_1^\nu J_{m\nu}K_{m\mu\alpha} \right) \delta \theta_1^\alpha= 0. 
\end{equation}
to second order, with the term 
$r_m K_{m\mu\alpha} \delta \theta_1^\alpha$
the previously neglected term.

We now turn our attention to the second term in parentheses in Eq.~(\ref{LM:eq:normal2}).  Using the definition of $\delta \theta_1 = -(J^TJ + \lambda D^TD)^{-1} J^T r$, we can write
\begin{eqnarray}
  r_m K_{m\mu\alpha}  + \delta \theta_1^\nu J_{m\nu}K_{m\mu\alpha}  & = &
  r_m K_{m\mu\alpha}  - r_m J_{m\beta} (J^TJ + \lambda D^TD)^{\beta \nu} J_{n\nu} K_{n\mu\alpha} \nonumber \\
\  & = & r_m \left( \delta_{mn} - J_{m\beta} (J^TJ + \lambda D^TD)^{\beta \nu} J_{n\nu} \right) K_{n\mu\alpha}. \nonumber
\end{eqnarray}
We now make an appeal to geometric considerations by noting that $\delta_{mn} - J_{m\beta} (J^TJ + \lambda D^TD)^{\beta \nu} J_{n\nu} = P^N_{mn}$ is a matrix that projects vectors perpendicular to the tangent plane of the \emph{Model Graph} as described in reference\cite{Transtrum2011}.  If the curvature of the model graph is small, then $P^N K \approx 0$ and this term can be neglected.  Furthermore, as the algorithm nears the best fit, the residuals are very nearly orthogonal to the tangent plane (regardless of whether $\vert r \vert$ is small), so that $P^N r \approx 0$ also.  We therefore assume that this term is negligible compared to other corrections.

Returning to Eq.~(\ref{LM:eq:normal2}), after ignoring the last term in parentheses, we find
\begin{eqnarray}
  \delta \theta_2 & = & -1/2\, \left( J^TJ + r^TK + \lambda D^TD \right)^{-1} J^T r'' \nonumber \\
  \ & \approx &  -1/2\, \left( J^TJ + \lambda D^TD \right)^{-1} J^T r'', \nonumber
\end{eqnarray}
where we have used the directional second derivative $r_m'' = K_{m\mu\nu} \delta \theta_1^\mu \delta \theta_1^\nu$ and in the second line made the usual approximation to the Hessian, giving the formula originally presented in \cite{Transtrum2010}.  This formula was originally interpreted as the second order correction to geodesic flow on the model graph, and so we refer to this correction as the geodesic acceleration correction.  By analogy, we refer to the first order correction as the velocity:
\begin{equation}
  \label{LM:eq:veldef}
  \delta \theta = \delta \theta_1 + \delta \theta_2 \equiv v \delta t + 1/2\, a \delta t^2.
\end{equation}

It is surprising how good the small-curvature approximation turns out to be.  As shown in reference \cite{Transtrum2011}, the extrinsic curvature of the model manifold should under many circumstances be very small compared the largest step size an algorithm can take, which is limited by the so-called parameter-effect curvature.  When Bates and Watts first introduced measures of parameter-effects curvature, they noted that for every problem they considered the parameter-effects curvature was larger (often much larger) than the extrinsic curvature\cite{Bates1980}; explicit examples in \cite{Transtrum2011} show parameter-effects curvature with up to six order of magnitude larger extrinsic radii of curvature compared to the allowed step sizes.  Although there are assuredly counter examples, it is reasonable to expect that for most problems of interest, this approximation will be excellent, and the geodesic acceleration will capture most of the next-order correction in the cost.

The geodesic acceleration depends on the second derivative of the model, but perhaps surprisingly, the only dependence is on the directional second derivative oriented along the first order correction, $\delta \theta_1$.  This result is significant, as calculating the full array of second derivatives would likely be prohibitively expensive for large models.  However, a directional second derivative has a small computational cost, comparable to a single evaluation of $r(\theta)$, and in fact a finite-difference estimate can be found with only one additional evaluation of $r$.  In contrast, for large models, most of the computational cost of minimizing least-squares problems comes from evaluating the Jacobian matrix.  In these cases, the additional cost of evaluating the second order correction is negligible.

\begin{figure}[htbp]
  \centering
  \includegraphics[width=6in]{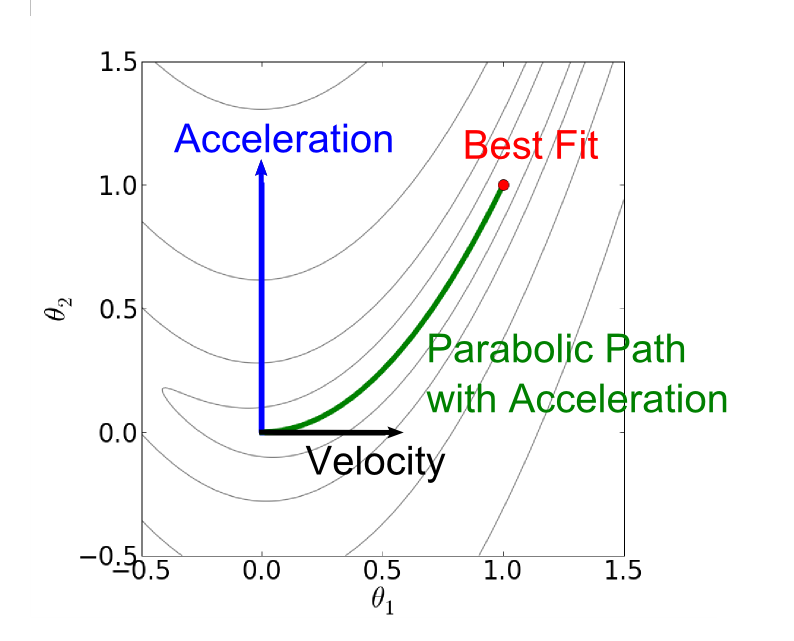}
  \caption[Geodesic acceleration in a canyon]{When navigating a canyon toward the best fit, the geodesic acceleration indicates in which direction the canyon is curving.  By approximating the path with a parabola, the best fit can be found in fewer iterations.}
  \label{LM:fig:acceleration}
\end{figure}

One benefit of including the geodesic acceleration comes when the algorithm is navigating a narrow canyon toward the best fit as illustrated in Figure \ref{LM:fig:acceleration}.  By approximating the path with a parabola, the geodesic acceleration method can more accurately follow the path of a winding canyon toward the best fit.

In order to utilize the geodesic acceleration as an addition to the Levenberg-Marquardt algorithm, it is necessary to make one other small addition.  We require acceptable steps to satisfy the condition
\begin{equation}
  \label{LM:eq:va}
  \frac{ 2 \vert \delta \theta_2 \vert} {\vert \delta \theta_1 \vert} \leq \alpha
\end{equation}
where $\alpha$ is some number of order $1$ that is set by the user and whose optimal value may vary from problem to problem.  The motivation for this requirement is that the proposed step represents a truncated perturbation series and so the terms ought to be decreasing in magnitude to guarantee convergence. For most problem we find that $\alpha = 0.75$ is a good guess.  Problems for which convergence is difficult, $\alpha = 0.1$ is an effective choice.

We note that for a given value of $\alpha$ one can always find a suitable value of $\lambda$ such that Eq.~(\ref{LM:eq:va}) is satisfied as long as the second derivative is reasonably well behaved.  In particular, if $\lambda$ is very large, then $(J^TJ + \lambda D^TD)^{-1},  \approx \frac{1}{\lambda} (D^TD)^{-1}$, and
\begin{equation}
  \label{LM:eq:thetadotlargelam}
  \delta \theta_1 \approx -\frac{1}{\lambda} (D^TD)^{-1} \nabla C.
\end{equation}
We define $u = \lambda \delta \theta_1$ which is (asymptotically) independent of $\lambda$, so that $r_m'' = K_{m\mu\nu} \delta \theta_1^\mu \delta \theta_1^\nu \approx \frac{1}{\lambda^2} K_{m\mu\nu} u^\mu u^\nu$.  Then it follows that
\begin{equation}
  \label{LM:eq:thetadotdotlargelam}
  2 \delta \theta_2 \approx \frac{1}{\lambda^3} \left( D^T D \right)^{-1} J^T \left( u^T K u \right).
\end{equation}
Therefore, as long as $u^T K u$ is bounded, Eq.~(\ref{LM:eq:va}) can be satisfied by selecting a sufficiently large value of $\lambda$.

\begin{figure}[htbp]
  \centering
  \includegraphics[width=6in]{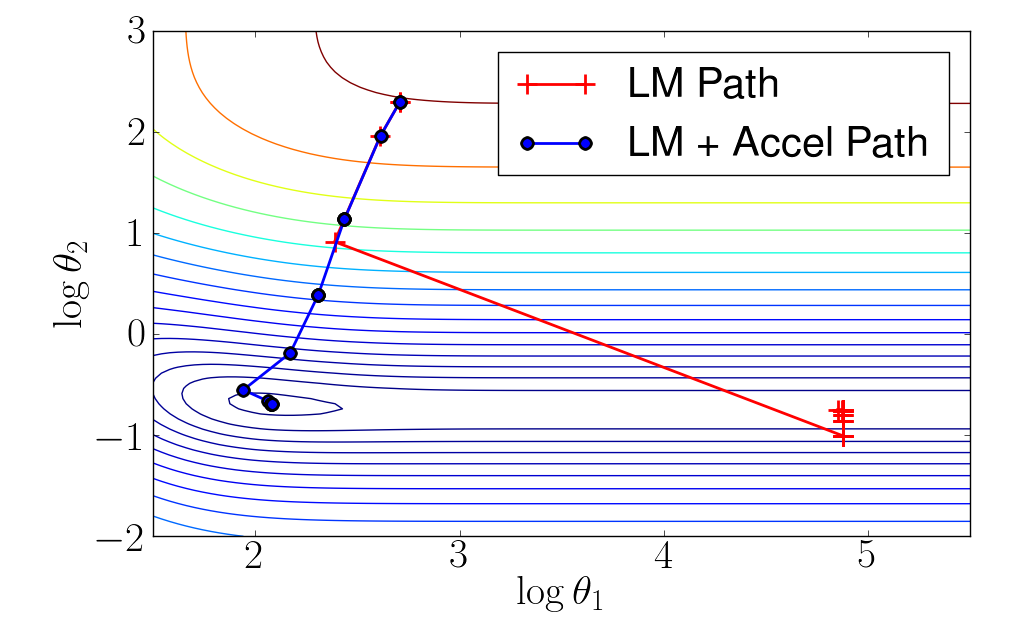}
  \caption[Geodesic Acceleration helps avoid Parameter Evaporation]{Here we minimize the problem in figure \ref{LM:fig:Contours} in $\log$ parameters.  When the algorithm starts on a plateau, it can take large, uncontrolled steps, that although they decrease the cost, they may move the algorithm farther from the best fit. In this example the standard algorithm without acceleration becomes stuck in a very flat region where $\nabla C \approx 0$ within numerical precision; however, this point is a poor fit to the data and not a local minimum.  By including acceleration and enforcing Eq.~(\ref{LM:eq:va}), the algorithm recognizes that these steps are too large to be trusted since $\vert \delta \theta_2 \vert > \vert \delta \theta_1 \vert$.  It therefore takes smaller steps, which allow it to ultimately find the best fit.}
  \label{LM:fig:ParameterEvaporation}
\end{figure}

Without the requirement in Eq.~(\ref{LM:eq:va}), the resulting algorithm may be unpredictable and will often take large, uncontrolled steps and become lost.  This phenomenon is closely related to parameter evaporation that gives so much trouble to the standard algorithm. The additional requirement in Eq.~(\ref{LM:eq:va}) helps the geodesic algorithm avoid parameter evaporation, increasing its likelihood of convergence.  In figure \ref{LM:fig:ParameterEvaporation} we show that adding geodesic acceleration with the requirement in Eq.~(\ref{LM:eq:va}) can improve the behavior of teh standard Levenberg-Marquardt algorithm.

In section \ref{LM:sec:DTD} it was noted that for a suitable choice of $D^TD$ the iterates produced by the Levenberg-Marquardt algorithm are invariant to a change of scale of the parameters.  It is interesting to note that this result is unchanged by including the geodesic acceleration.  It is also important to note that by including the geodesic acceleration, the step size is no longer a monotonically decreasing function of $\lambda$.  Furthermore, including the geodesic acceleration makes it computationally unreasonable to calculate $\lambda$ for a specified step size $\Delta$.  Rather, in our implementation, we implement indirect methods by selecting $\lambda$ so that $\vert \delta \theta_1 \vert = \Delta$.  With this convention, the Levenberg-Marquardt method with geodesic acceleration is a type of hybrid method between a trust region and line search method.  In particular, the choice of damping, $\lambda$, can be considered a trust region in $\delta \theta_1$.  Having selected the first order correction $\delta \theta_1$, the second order correction explores the model behavior in the chosen direction and diverts and dampens the step as appropriate.

Before discussing the performance of the geodesic acceleration, we offer a few remarks about evaluating the directional second derivative.  As is always the case, analytic derivatives are preferable to finite difference estimates.  If an analytic expression can be found for the first derivatives, in principle one could also find an expression for the directional second derivative, although the resulting expression may be very complicated, especially for large models.  (With code often being generated by computer algebra systems, it may not be unreasonable in some cases.)  Automatic differentiation also might make exact evaluations of the second derivative feasible.

In cases where a directional second derivative cannot be evaluated analytically, one can always use a finite difference approximation.  The usual formula for the directional second derivative is
\begin{equation}
  \label{LM:eq:AvvFD2}
  K_{m\mu\nu} \delta \theta^\mu \delta \theta^\nu \approx \frac{r_m(\theta + h \delta \theta_1) - 2 r_m(\theta) + r_m(\theta - h \delta \theta_1) }{h^2},
\end{equation}
for some finite-difference step size, $h$.  When evaluating Eq.~(\ref{LM:eq:AvvFD2}), note that the algorithm will already have evaluated $r(\theta)$, leaving two additional function evaluations necessary for the estimate.  The algorithm has also previously evaluated the Jacobian matrix.  Using this information one can find a finite difference estimate with just a single additional function evaluation:
\begin{equation}
  \label{LM:eq:AvvFD1}
  K_{m\mu\nu} \delta \theta^\mu \delta \theta^\nu  \approx \frac{2}{h} \left( \frac{r_m(\theta + h \delta \theta_1) - r_m(\theta)}{h} - J_{m\mu}\delta \theta_1^\mu   \right).
\end{equation}
In practice, the finite difference estimate may be sensitive to $h$, particularly if the problem is poorly scaled.  We find in practice that choosing a large finite-difference step size, giving something analogous to a secant estimation, is less dangerous than a step that is too small. In general, choosing $h = 0.1$, so the step is about $10\%$ of the first order step seems to work reasonably well.

We now consider how the geodesic acceleration affects performance, detailed results are presented in  \ref{LM:sec:accelcomparison}.  In our experiments, including geodesic acceleration correction improved the algorithm's ability to converge to good fits and to do so with less computational cost.  On some problems, the geodesic acceleration converged with an average of a factor of 70 fewer Jacobian evalautions!  Perhaps the most significant benefit gained from geodesic acceleration, however, is in the improved success rate and fit quality.  We attribute this improvement to the modified acceptance criterion given in Eq.~(\ref{LM:eq:va}).  For many cases the algorithm could be improved further by a smaller choice of $\alpha$ (the results in  \ref{LM:sec:accelcomparison} are for $\alpha = 0.75$), although this comes at a cost in convergence speed (more Jacobian evaluations).

\section{Uphill steps}
\label{LM:sec:uphill}

It is necessary for the algorithm to have some way to distinguish whether a proposed step should be accepted or rejected.  The standard choice in the Levenberg-Marquardt method is to accept all steps that decrease the cost and reject all steps that increase the cost.  Although this is a natural and safe choice, it is often not the most efficient, as we demonstrate in this section.

When an algorithm must follow a narrow canyon to the best fit, if the aspect ratio of the canyon is very large, then there will be only a small sliver of steps that decrease the cost.
Uphill moves, akin to the path followed by a bobsled racer, allow more rapid progress toward the best fit.
Accordingly we modify the acceptance criterion following a method proposed by Umrigar and Nightingale\cite{UmrigarBold}, such that downhill moves are always accepted, but uphill moves only
conditionally.  To determine whether an uphill move is accepted at
each iteration $i$, we compute
\begin{equation}
\beta_i=\cos(\delta \theta^\textrm{new}_1, \delta \theta_1^\textrm{old}),
\label{eq.cos-on}
\end{equation}
which denotes the cosine of the angle between the proposed velocity $\delta \theta_1^\textrm{new}$ and the velocity of the last accepted step $\delta \theta_1^\textrm{old}$.
The idea is to accept uphill moves if the angle
$\mbox{arccos}\,\beta_i$ is acute, with increasing acceptance as the
angle more nearly vanishes.
To be specific, we accept an uphill move if
\begin{equation}
(1-\beta_i)^b\,C_{i+1} \le  C_i,
\label{eq.boldness1}
\end{equation}
or more conservatively if
\begin{equation}
  (1-\beta_i)^b\,C_{i+1} \le  \min(C_1,\dots,C_i),
\label{eq.boldness2}
\end{equation}
where $\min(C_1,\dots,C_i)$ is the smallest cost yet found.
Reasonable choices for the value of $b$ are 1 or 2, with $b=2$ allowing higher uphill moves than $b=1$.  In combination
with either one of the Eqs.~(\ref{eq.boldness1}) or
(\ref{eq.boldness2}), this yields four possible variants of the algorithm.  This method for accepting steps was originally developed by Umrigar and Nightingale\cite{UmrigarBold} in 1994 and has been employed by them for optimizing many-body wave functions used in quantum Monte Carlo calculations.  We refer to this method of accepting steps as the ``bold'' acceptance criterion.

In addition to frequently reaching minima in a smaller number of optimization steps, we find that on average better minima are found,
for systems with multiple minima, when uphill moves are allowed.  The reason is that by not following the valley floor closely
the optimization avoids getting trapped in some of the potholes it encounters along the way.

We now investigate how the bold acceptance criterion affects the algorithms' performance, with detailed comparisons given in  \ref{LM:sec:boldcomparison}.  While the results vary from problem to problem, it appears that for many cases the bold acceptance can speed up the performance of the algorithm, in some cases by a factor of $30$.  Unfortunately, this increase in speed comes at a cost in the stability of the algorithm; in many cases the success rate of the algorithm drops when the bold method is used.  This should be expected, however.  If the standard algorithm is prone to get lost on flat regions of the cost surface by taking uncontrolled steps, then allowing the method to move uphill should produce even less constrained steps.  

The effect of lower success rates when using bold acceptance can be partially mitigated by including the geodesic acceleration correction and enforcing Eq.~(\ref{LM:eq:va}).  By reducing the value of $\alpha$ the algorithm can usually be made more successful at the expense of a slower algorithm.  The trade-off between a stable algorithm and a fast algorithm seems to be an inherent feature of this problem. 

We note that in our experiments, we have used relatively difficult starting points on many of the test problems.  If we had used the standard starting points supplied by the Minpack-2 or NIST versions of problems A - N, the success rate and fit quality would have been much higher.  This result is a reflection of that fact that there is no guarantee for convergence when uphill moves are accepted.  On the other hand, in our experience if a problem is chosen such that there is little chance of the algorithm becoming lost or not converging, then accepting uphill moves can greatly reduce the time to find good fits.  This is most likely to be useful for fitting problems that either start in a narrow canyon or can easily find the canyon but become sluggish en route to the best fit.  If the problem is difficult because it is hard to find a canyon, then accepting uphill moves are not likely to improve the search.

\section{Updating the Jacobian Matrix}
\label{LM:sec:Broyden}

When comparing algorithm performance, we have assumed that the most computationally intensive part of the Levenberg-Marquardt algorithm is an evaluation of the Jacobian matrix of first derivatives.  If this is done using finite difference then for a model of $N$ parameters the Jacobian matrix is $N$ times as expensive as a residual evaluation.  If analytic formulas are available, it may be more efficient than a finite difference approximation, but for large $N$ Jacobians will still occupy the bulk of the computer time.  Much of the discussion of this paper has revolved around reducing the number of Jacobian evaluations necessary for convergence.

Typically, the algorithm will evaluate the Jacobian after each accepted step in order to calculate the gradient $\nabla C = J^T r$ and the matrix $g = J^T J + \lambda D^T D$.  Broyden suggested a quasi-Newton root finding method that evaluates the Jacobian only on the first iteration and subsequently updates the Jacobian with a rank-$1$ update\cite{Broyden1965}.  Thus, given the Jacobian at the previous iterate, $J_{i-1}$, the Jacobian at the current parameter values $J_i$ is estimated to be
\begin{equation}
  \label{LM:eq:Broyden}
  J_i=J_{i-1} + \frac{\Delta r_i-J_{i-1}\Delta \theta_i}{|\Delta \theta_i|^2}\Delta \theta_i^T,
\end{equation}
where $\Delta r_i = r_i - r_{i-1}$ is the change in the residual vector and $\Delta \theta_i = \theta_i - \theta_{i-1}$ is the change in the parameter space vector between iterations.  In principle this method can be applied to the Levenberg-Marquardt method to eliminate the need to evaluate $J$ at each step.

In practice, after many such updates, the matrix $J$ may become a very poor approximation to the actual Jacobian, resulting in a poor estimate of the gradient direction.  If the algorithm's performance suffers as a result, it may be necessary to reevaluate the whole Jacobian matrix and begin the update scheme anew.  We therefore reevaluate the Jacobian after a few proposed steps have been rejected by the algorithm.  Typically, reevaluating after one or two rejections works well.

When the geodesic acceleration is included, then each iteration has the information of two function evaluations and one can construct a rank-$2$ update to the Jacobian.  We accomplish this as follows: if a step is proposed with velocity $\delta \theta_1 = v$ and acceleration $\delta \theta_2 = a/2$, then we first assume a step was taken corresponding to $\delta \theta = v/2$ and residuals $ r_i + \frac{1}{2}J_i v + \frac{1}{8} v^T  K_i v$.  Notice this expression for the residuals involves the directional second derivative which has been evaluated to calculate the geodesic acceleration.  A rank-$1$ Broyden update is then performed for this step.  It is then assumed that a second step is taken corresponding to $\delta \theta = \frac{1}{2} \left( v + a \right) $ and another rank-$1$ update is performed corresponding to this step.

In practice there is little performance gain from using the rank-$2$ update instead of the rank-$1$ update suggested by Broyden.  This is most likely because the velocity and the acceleration are often nearly collinear, as described in \cite{Transtrum2011}.  However, as there are cases in which a rank-$2$ update can be beneficial, we include it in our implementation.

The details of our tests using rank-deficient updates is given in  \ref{LM:sec:boydencomparison}.  We find that, when successful, the Broyden update scheme can dramatically speedup the algorithm, requiring many fewer evaluations of the Jacobian matrix.  However, the algorithm also appears to be more likely to get lost.  We understand the lower success rates and fit quality as a direct consequence of using an approximate Jacobian matrix, resulting in an inaccurate gradient direction.  

By using a rank-$2$ update with the geodesic acceleration, the success rate of the algorithm is not improved significantly.  It is likely that a more sophisticated scheme for deciding when the Jacobian should be reevaluated could improve the method's robustness against becoming lost.  We recommend using this Jacobian update scheme only when there is very little chance that the algorithm will become lost or when the Jacobian is so expensive that evaluating it multiple times is not possible.

Particularly for large problems, the relative cost of a Jacobian evaluation to a function evaluation reflects the relative information content of the two.  For small problems the rank-$1$ update contains a significant fraction of the information  available in the Jacobian.  However, for larger problems the Broyden update becomes an increasingly bad approximation.  In these cases, the authors speculate significant performance gains could be obtained by updating the Jacobian with a few strategically chosen function evaluations or directional derivatives rather than updating the entire Jacobian.

\section{Conclusion}
\label{LM:sec:conclusion}

The computational problem of minimizing a sum of squares is a common problem, particularly in data fitting, that can be notoriously difficult.  The difficulties often fall into one of two categories: algorithms easily become lost on broad flat plateaus or become sluggish as they must follow narrow canyons to the best fit.  In this paper we have discussed several modifications to the standard least-squares minimizer, the Levenberg-Marquardt algorithm, to help address these difficulties.

We have derived the ``geodesic acceleration'' correction to the Levenberg-Marquardt step, by including second order corrections in the Taylor approximation of the residuals and assuming that the extrinsic curvature on the model graph as described in reference \cite{Transtrum2010, Transtrum2011} is small.  This correction can be computed at each step with minimal additional computational cost. We have find that this correction helps the algorithm be both more robust to initial conditions, resulting in higher success rates, as well has decreasing the computational cost of finding the best fit as measured by the number of Jacobian evaluations.  

We have also suggested that accepting uphill steps in a controlled manner can also speed up the algorithm.  When an algorithm is susceptible to becoming lost in parameter space, accepting uphill moves may exacerbate this problem, but when the algorithm must follow a narrow canyon to the best fit, the potential speed up of the bold method can be enormous.  We have also suggested that providing a rank-deficient update to the Jacobian matrix can further reduce the computational cost of the Levenberg-Marquardt algorithm.  Although the resulting algorithm has a tendency to become lost, it can be much more efficient when following a canyon to the best fit.

\begin{figure}[htbp]
  \centering
  \includegraphics[height=7in]{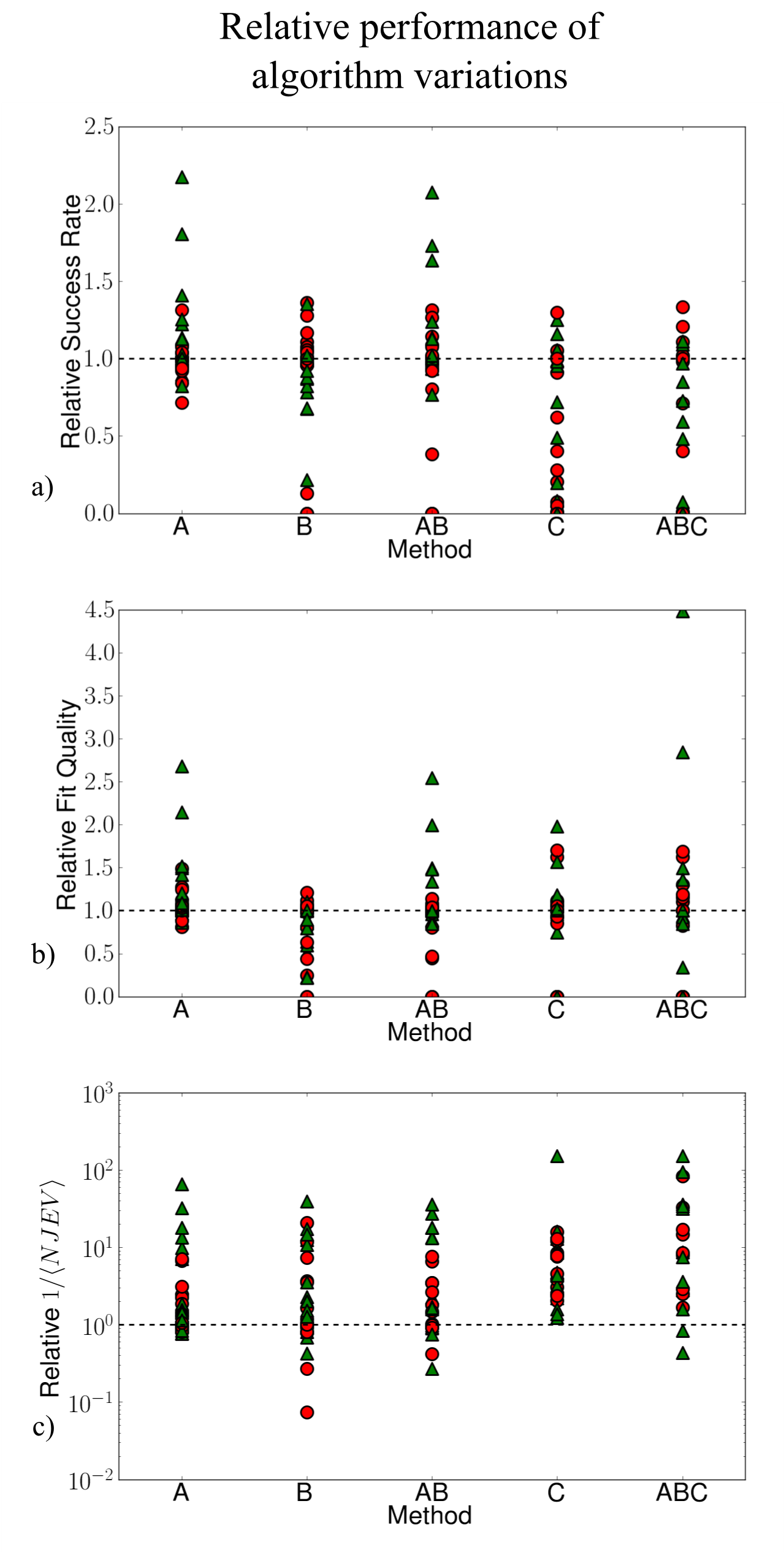}
  \caption[Summary of algorithms' performances]{\textbf{Performance of several algorithms}  The relative success rate (a), fit quality (b), and inverse NJEV (c) of several algorithms.  The rates are each relative to the same algorithm without geodesic acceleration, bold, or a Broyden update.  Columns represent different algorithms, with red dots denoting the $\lambda 2 / 3$ method while green triangles represent the $\Delta 2 / 3$ method on each of the $17$ test problems.  The label A indicates that geodesic acceleration was including, B denotes uphill steps were accepted with the bold acceptance criterion, and C indicates that the Jacobian was updated with Broyden's method. }
  \label{LM:fig:Summary}
\end{figure}

The performance of our several suggested improvements is summarized in figure \ref{LM:fig:Summary}.  Notice that including geodesic acceleration has the tendency to improve success, fit quality, and speed (see  \ref{LM:sec:successmeasures} for a definition of these measurements).  Including the bold acceptance and using Broyden's update can be even more effective for speeding up the algorithm, although on some problems, particularly those with difficult start points, these algorithms are more likely to become lost.  Which variation of the algorithm is most effective is likely to vary from problem to problem and whether the user is more concerned with fit quality or convergence speed.  A good implementation of the Levenberg-Marquardt method should be flexible enough to allow the user to adapt the method to their specific needs.

We have provided an open source (FORTRAN) version of the Levenberg-Marquardt algorithm together with our suggested improvements\cite{SFGeoLM}.  Our implementation provides a simple way for each addition to be turned on or off, in addition to choosing among several schemes for updating the damping term $\lambda$.  In this way, users have the tools to optimize the fitting process to more quickly and robustly find best fits based upon the needs of their individual models.

\section*{Acknowledgments}

The authors would like to thank C.~Umrigar and P.~Nightingale who originally conceived of the bold acceptance method and suggested it to us in private conversation.  The authors would also like to thank C.~Umrigar, P.~Nightingale, B.~Machta, and S. Teukolsky for helpful conversations.  This work was supported by NSF grant number DMR-1005479.

\appendix

\section{Test Problems}
\label{LM:sec:testprobs}

In order to gauge the relative effectiveness of the improvements described in this paper, we use $17$ test problems (denoted by the letters A-Q throughout this work) which we take from the Minpack-2 project \cite{Averick1992} and the NIST Statistical Reference datasets \cite{mccullough1998} and some of our own research and that of Umrigar and Nightingale.  We summarize these problems in this appendix.

Although we use several standard test problems, we emphasize that we do not use the standard starting points for these problems.  For each of these problem, we choose starting points from a Gaussian distribution centered at one of the suggested starting points for each problem.  The width of the Gaussian is manually adjusted until one of the standard algorithms begins to show a noticeable variation in performance among the points.  This method of choosing the starting points makes many of the test problems much harder than they otherwise would have been.  It is fortunate that by choosing starting points in this way the easy problems can be made of comparable difficulty to the more realistic problems.  In particular, the ease and quick evaluation of the smaller problems make them ideal test cases provided they can be made sufficiently difficult to imitate more realistic problems.  In addition to making the problems more difficult, by considering the performance from several starting points, we can avoid the complication that an algorithm may perform well by accident for a particular starting point.  We provide the specific starting points for each algorithm in addition to source code that runs each problem\cite{SFGeoLM}.

Problem A: Isomerization of $\alpha$-pinene (Direct formulation) taken from the Minpack-2 project, consisting of five parameters and $40$ residuals.  This model is evaluated as a linear ordinary differential equation with unknown coefficients.

Problem B: Isomerization of $\alpha$-pinene (Collocation formulation) taken from the Minpack-2 project, consisting of $130$ parameters and $165$ residuals.  This is an example of a constrained optimization problem in which the constraint is implemented as an $l_2$ penalty.  In our implementation, we have used the relatively weak penalty strength of $\sigma = 1000$ (as opposed to $\sigma = 10^6$ as suggested in \cite{Averick1992}).  As the strength of $\sigma$ is increased, the algorithm must more closely maintain the constraints at each iteration, making the algorithm become much slower.  Anecdotally, we observe that geodesic acceleration and bold acceptance can be very helpful in these cases.

Problem C: Coating thickness standardization taken from the Minpack-2 project, consisting of $134$ parameters and $252$ residuals.  This problem is a multiple-response data-fitting problem.  Because of its larger size it is one of the more computationally intensive problems in the set.

Problem D: Exponential data fitting taken from the Minpack-2 Project, consisting of $5$ parameters and $33$ residuals.  The functional form of this problem is
\begin{equation}
  \label{LM:eq:edfform}
  y(t,\theta) = \theta_1 + \theta_2 e^{-t \theta_4} + \theta_3 + e^{-t \theta_5}.
\end{equation}
This problem similar to those used in references \cite{Transtrum2010,Transtrum2011} for which geodesic acceleration was shown to be very effective.

Problem E: Gaussian data fitting taken from the Minpack-2 Project, consisting of $11$ parameters and $65$ residuals.  The functional form is
\begin{equation}
  \label{LM:eq:gdfform}
  y(t,\theta) = \theta_1 e^{- \theta_5} + \theta_2 e^{-(t-\theta_9)^2 \theta_6} + \theta_3 e^{-(t-\theta_{10})^2 \theta_7} + \theta_4 e^{-(t-\theta_{11})^2 \theta_8}.
\end{equation}
This problem is difficult for starting points far from the minimum.

Problem F: Analysis of thermistor resistance taken from the Minpack-2 Project, also known as the MGH10 problem from the NIST dataset.  This problem consists of $3$ parameters and $16$ data points. The functional form of this problem is
\begin{equation}
  \label{LM:eq:atrform}
  y(t) = \theta_1 e^{\frac{\theta_2}{t-\theta_3}}.
\end{equation}
For starting points with large values of $\theta_3$, this problem becomes very difficult as the $t$ dependence is lost.  Including a small value of $\alpha$ in the geodesic acceleration acceptance criterion is very helpful to force the algorithm move toward smaller $\theta_3$ in this case.

Problem G: Analysis of enzyme reaction taken from the Minpack-2 Project, also known as the MGH09 problem from the NIST dataset.  This problem consists of $4$ parameters and $11$ data points.  This problem takes the form
\begin{equation}
  \label{LM:eq:aerform}
  y(t,\theta) = \frac{\theta_1 \left( t^2 + t \theta_2 \right) }{t^2 + t \theta_3 + \theta_4}.
\end{equation}
Many algorithms have a low success rate because of a local minimizer at infinity.  As discussed in \cite{Transtrum2011}, this scenario is likely to be a generic feature of large, ill-conditioned data fitting problems.

Problem H: Chebyshev quadrature taken from the Minpack-2 Project, consisting of $8$ parameters and $11$ residuals.
This problem exhibits a disparity between the success rate and the fit quality due to algorithms converging to local minima.

Problem I: Thurber problem from the NIST dataset, consisting of $7$ parameters and $37$ residuals.  This problem is a rational function of the form
\begin{equation}
  \label{LM:eq:Thurberform}
  y(t,\theta) = \frac{\theta_1 + \theta_2 t + \theta_3 t^2 + \theta_4 t^3}{1 + \theta_5 t + \theta_6 t^2 + \theta_7 t^3}
\end{equation}

Problem J: BoxBOD problem from the NIST dataset, consisting of $2$ parameters and $6$ residuals.  The functional form of the problem is
\begin{equation}
  \label{LM:eq:BoxBODform}
  y(t,\theta) = \theta_1 \left( 1- e^{-\theta_2 t} \right)
\end{equation}

Problem K: Rat42 problem from the NIST dataset, consisting of $3$ parameters and $9$ residuals.  The functional form of this problem is
\begin{equation}
  \label{LM:eq:Rat42form}
  y(t,\theta) = \frac{\theta_1}{1 + e^{\theta_2 - \theta_3 t}}
\end{equation}

Problem L: Eckerle4 problem from the NIST dataset, consisting of $3$ parameters $35$ residuals.  The functional form of this problem is
\begin{equation}
  \label{LM:eq:Eckerle4form}
  y(t,\theta) = \frac{\theta_1}{\theta_2} e^{\frac{-(t-\theta_3)^2}{2\theta_2^2}}
\end{equation}

Problem M: Rat43 problem from the NIST dataset, consisting of $4$ parameters and $15$ residuals.  The functional form of this problem is
\begin{equation}
  \label{LM:eq:Rat43form}
  y(t,\theta) = \frac{\theta_1}{\left( 1 + e^{\theta_2 - \theta_3 t} \right)^{1/\theta_4}}
\end{equation}

Problem N: Bennett5 problem from the NIST dataset, consisting of $3$ parameters and $154$ residuals.  The functional form of this problem is
\begin{equation}
  \label{LM:eq:Bennett5form}
  y(t,\theta) = \theta_1 \left( \theta_2 + t \right)^{-1/\theta_3}
\end{equation}

Problem O: A problem from systems biology described in \cite{Brown2004}.  This model consists of a differential equation model of $48$ parameters, mostly reaction rates and Michaelis-Menten constants fit to $68$ data points.  In order to help keep the parameters bounded, we have also introduced weak priors as described in \cite{Transtrum2011}.

Problem P: A problem for fitting a scaling function describing the distribution of avalanche sizes \cite{Chen2011}.  This model has $32$ parameters fit to $398$ data points.

Problem Q: A training problem for a feed forward artificial neural network.  The network is trained to data describing the compressive strength of concrete, as described in\cite{Yeh1998} and available here \cite{Frank+Asuncion:2010}.  In our formulation, there are $81$ parameters, consisting of the connection weights of the neural networks, and $1030$ data points.  We also include a weak quadratic prior on the parameters centered at zero in order to help avoid parameter evaporation.  These priors serve the same function as those in Problem O, described in \cite{Transtrum2011}.

Although there are other methods available for training artificial neural networks, they provide a good test problem for the general least-squares problem.  In particular, neural networks have many of the same properties as other fitting problems and provide an easy framework for varying the amount data, the number of inputs and outputs of the function as well as the number of parameters. They can also be evaluated relatively quickly and easily.

Problem R: A problem provided by Umrigar and Nightingale optimizing the Jastrow parameters of a variational wave function from Quantum Monte Carlo.  This problem consists of $27$ parameters and $2002$ residuals.

Again, the ensembles of initial parameter choices for each problem are provided\cite{SFGeoLM} to facilitate further comparisons to new algorithms.

\section{Detailed Comparison of Algorithm Performance}
\label{LM:sec:comparison}

\subsection{Measuring Success}
\label{LM:sec:successmeasures}

\begin{figure}[htbp]
\centering
\includegraphics[height=7in]{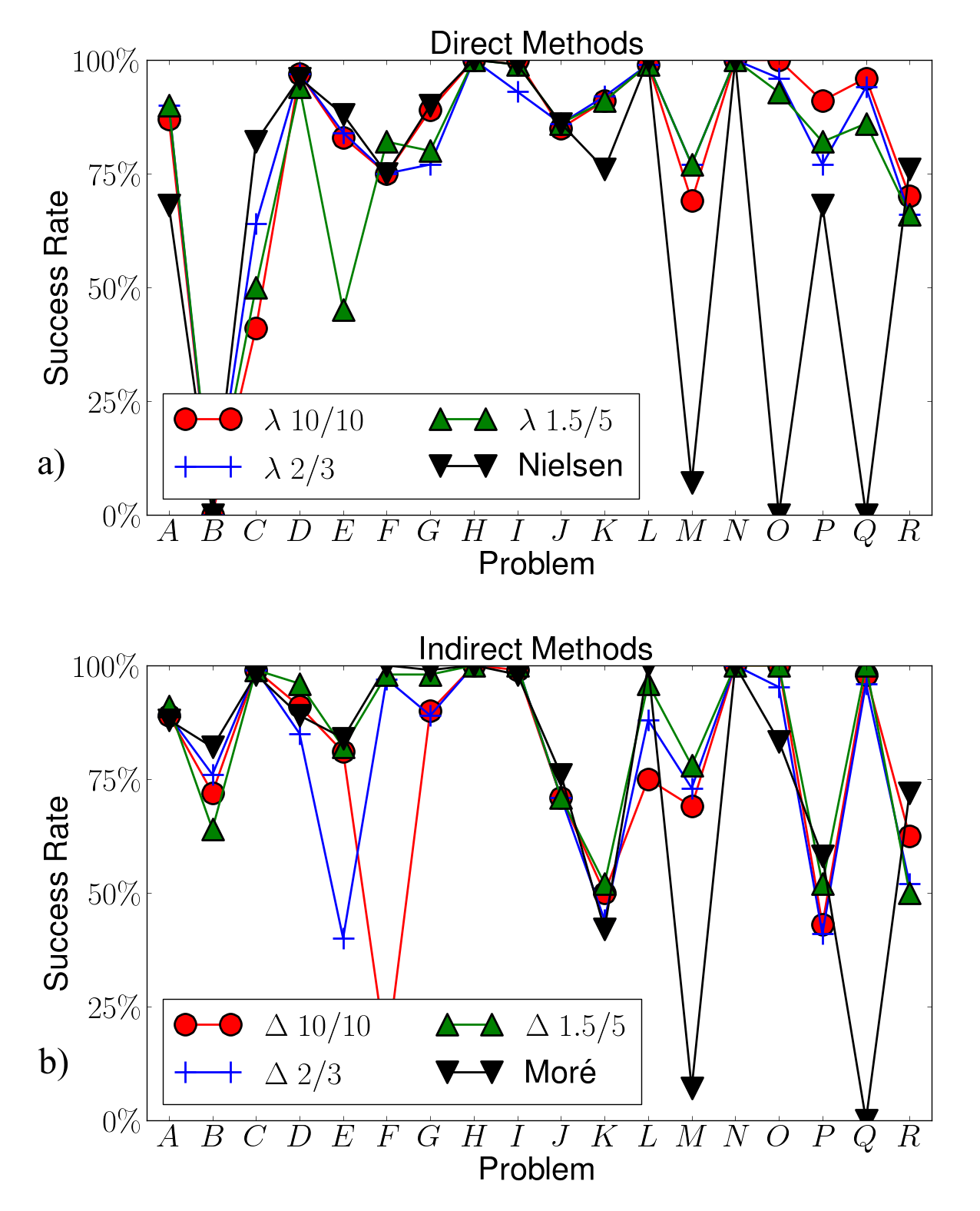}
\caption[Success rate of several algorithms]{ The average success rates for several direct (a) and indirect(b) methods on each of the $17$ test problems.}
\label{LM:fig:success}
\end{figure}

We give three measures of an algorithm's performance for each problem.  First, we consider the fraction of the attempts for which the algorithm claimed to have found a minimum; we refer to this as the success rate.  Since all the algorithms that we compare use the same convergence criterion described in section \ref{LM:sec:convergence}, this measure indicates to what extent the algorithm is able to avoid becoming lost in parameter space.  An algorithm with a high success rate was usually able to find a minimum within the alloted number of iterations.  In figure \ref{LM:fig:success} we plot the average success rate for several standard algorithms.

\begin{figure}[htbp]
  \centering
\includegraphics[height=7in]{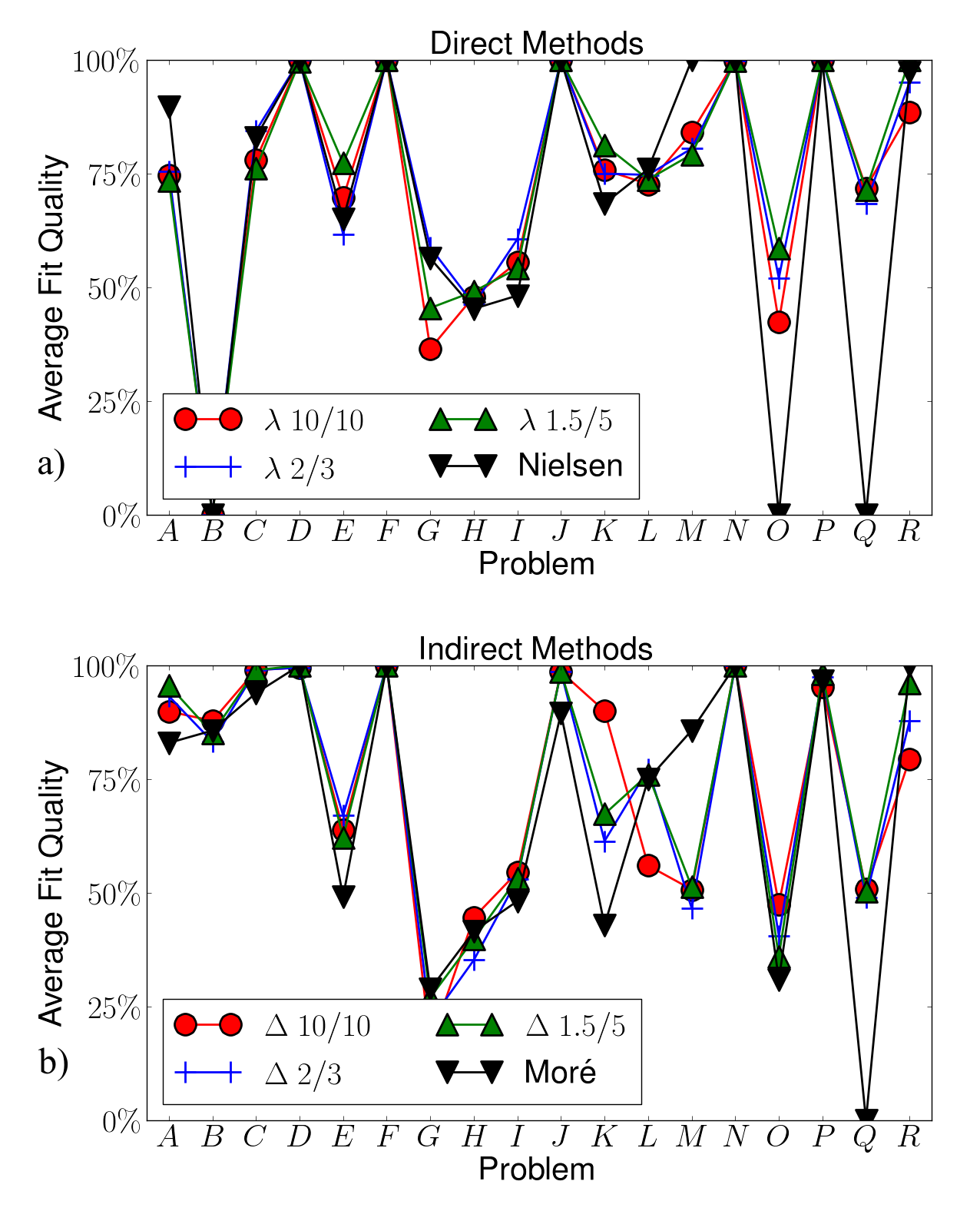}
\caption[Fit quality for several algorithms]{ The average quality factor, defined in Eq.~(\ref{LM:eq:Q}) for several direct (a) and indirect (b) methods on each of the seventeen test problems.}
\label{LM:fig:quality}
\end{figure}

Although an algorithm may claim success, it may have converged by quickly evaporating parameters and failed to have actually found a good fit.  To measure the relative quality of the fits, we define the factor
\begin{equation}
  \label{LM:eq:Q}
  Q = \exp\left(1-C_\textrm{final}/C_\textrm{best}\right),
\end{equation}
 where $C_\textrm{final}$ is the final cost found by the algorithm and $C_\textrm{best}$ is the best known cost.  (Although not applicable to any of these problem, a analogous formula for a problem whose solution has zero cost is $Q = \exp{-C_{\textrm{final}}/T}$, where $T$ is some tolerance.)  This term will be very near one if the algorithm has found the best fit, and exponentially suppressed otherwise.  For many of the problems from the Minpack-2 and NIST collections, the problems have either one minimum or a few minima with one much less than the others.  In these cases, Eq.~(\ref{LM:eq:Q}) will evaluate to either $0$ or $1$ depending on whether the best minima was found.  On the other hand, for many problems, particularly problems O and Q, algorithms will converge to a variety of local minima with a wide range of final costs.  In these cases, the quality factor, $Q$, will give partial weight to algorithms who find reasonable but not optimal fits.  Figure \ref{LM:fig:quality} displays the average value of this quality factor for each problem and several standard variations of the Levenberg-Marquardt algorithm.  Note that in calculating the average quality factor, we only include results for which an algorithm claimed success.

\begin{figure}[htbp]
\centering
\includegraphics[height=7in]{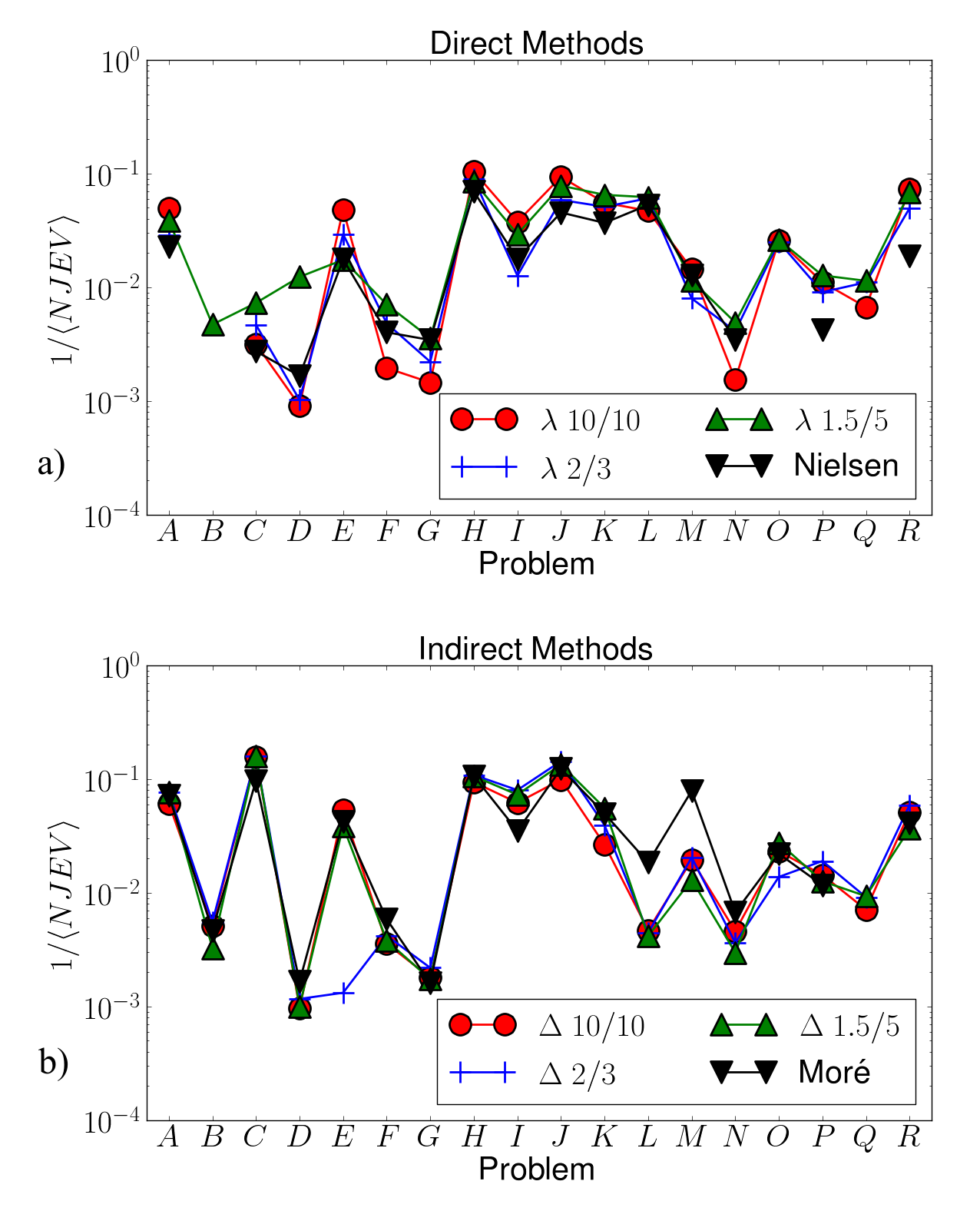}
\caption[Inverse NJEV for various algorithms]{ The inverse average Number of Jacobian Evaluations (NJEV) necessary for convergence for several direct (a) and indirect (b) methods on each of the seventeen test problems. }
\label{LM:fig:njev}
\end{figure}

Finally, in order to gauge the efficiency with which an algorithm converges to the best fit, we choose as a measure the number of Jacobian evaluations.  The advantage of this measure is that it is easy to extrapolate results for these simple test problems to larger, more computationally intensive problems where most of the computer time is spent calculating the Jacobian matrix.  Often, an algorithm will converge quickly to a poor fit.  In order to not bias results in favor of algorithms which find poor fits quickly, we calculate a weighted average of the number of Jacobian evaluations, weighted by the quality factor $Q$ in Eq.~(\ref{LM:eq:Q}).  As a convention, we plot the inverse of the average number of Jacobian evaluations so that larger numbers are preferable. The inverse average Number of Jacobian Evaluations (NJEV) for several standard algorithm variations  is shown in figure \ref{LM:fig:njev}.

\subsection{Comparison of Standard Methods}
\label{LM:sec:standardcomparison}

\begin{figure}[htbp]
  \centering
  \includegraphics[width=6in]{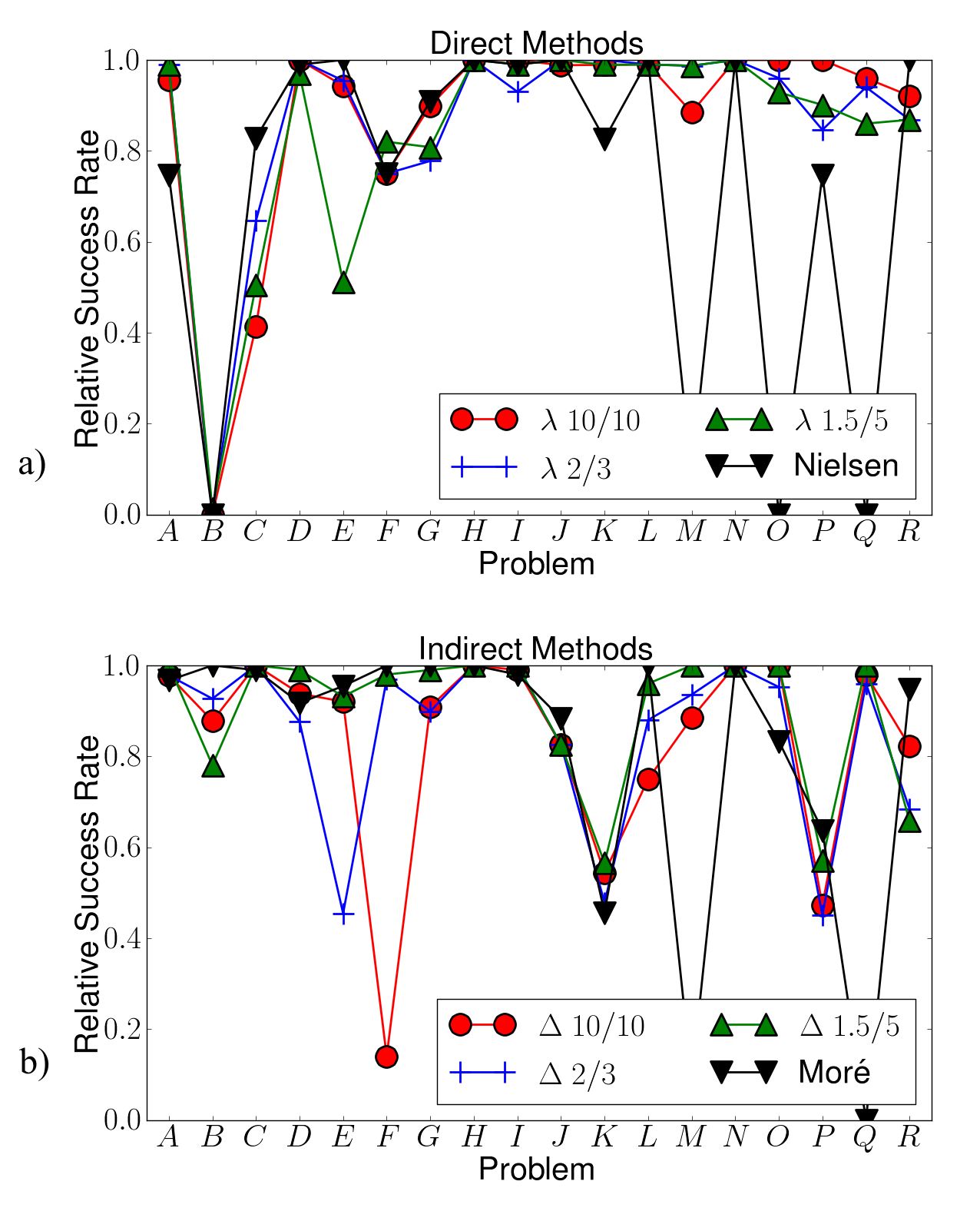}
  \caption[Relative success rates]{\textbf{Relative Success Rate} of direct (a) and indirect (b) methods of choosing $\lambda$.  The relative convergence rate is found by dividing each algorithm's convergence rate, as described in \ref{LM:sec:testprobs}, by the largest convergence rate of any method. Notice the direct methods' relative failure on problems B and C while the indirect methods struggled on J, K, and P.}
  \label{LM:fig:successrel}
\end{figure}

\begin{figure}[htbp]
  \centering
  \includegraphics[width=6in]{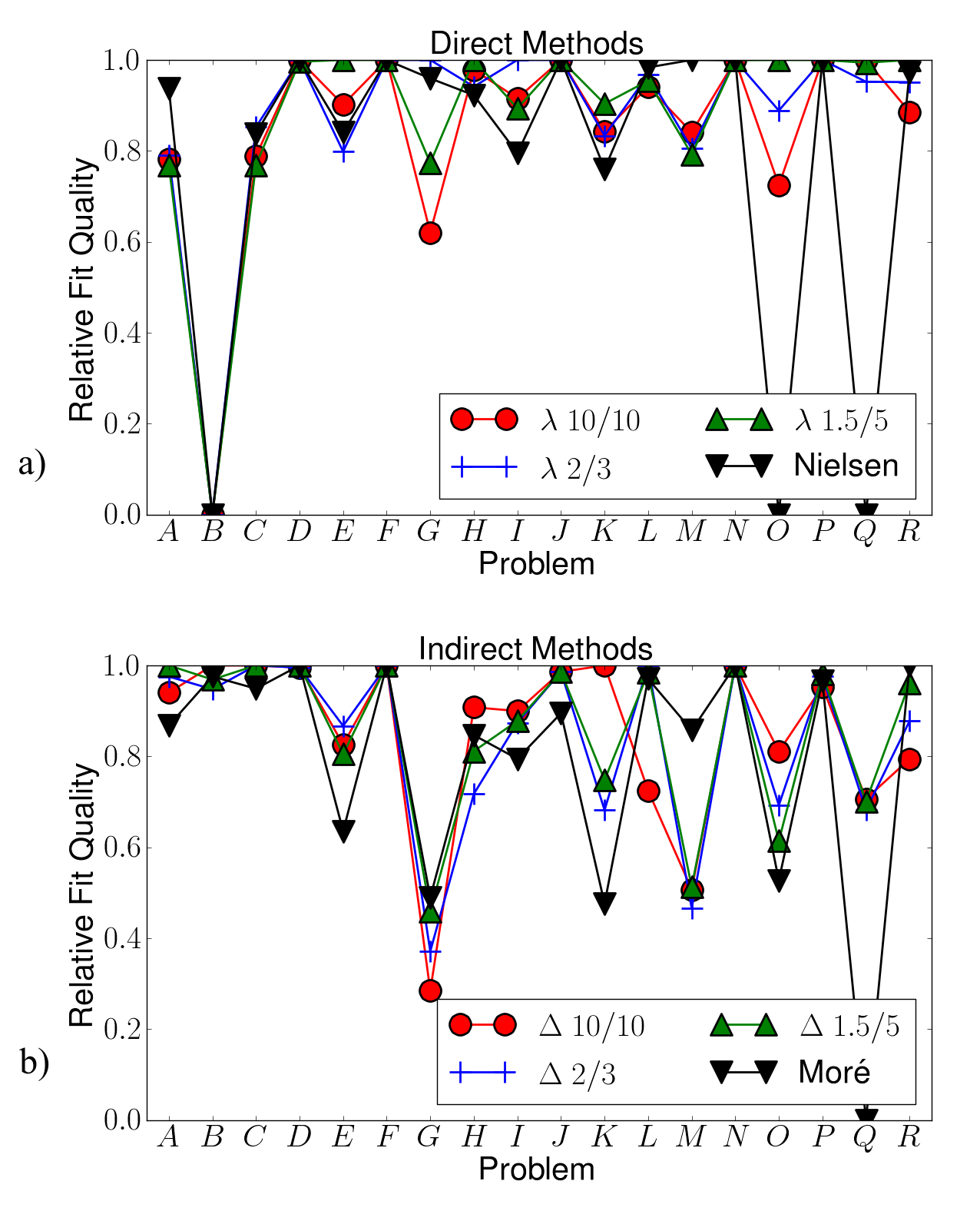}
  \caption[Relative Fit Quality]{\textbf{Relative Fit Quality} of direct (a) and indirect (b) methods of choosing $\lambda$.  The relative fit quality is found by dividing each algorithm's quality factor $Q$, as described in \ref{LM:sec:testprobs}, by the largest quality factor of any method.  Again note the poor fit quality of direct methods on problems B and C.  }
  \label{LM:fig:qualitysrel}
\end{figure}

\begin{figure}[htbp]
  \centering
  \includegraphics[width=6in]{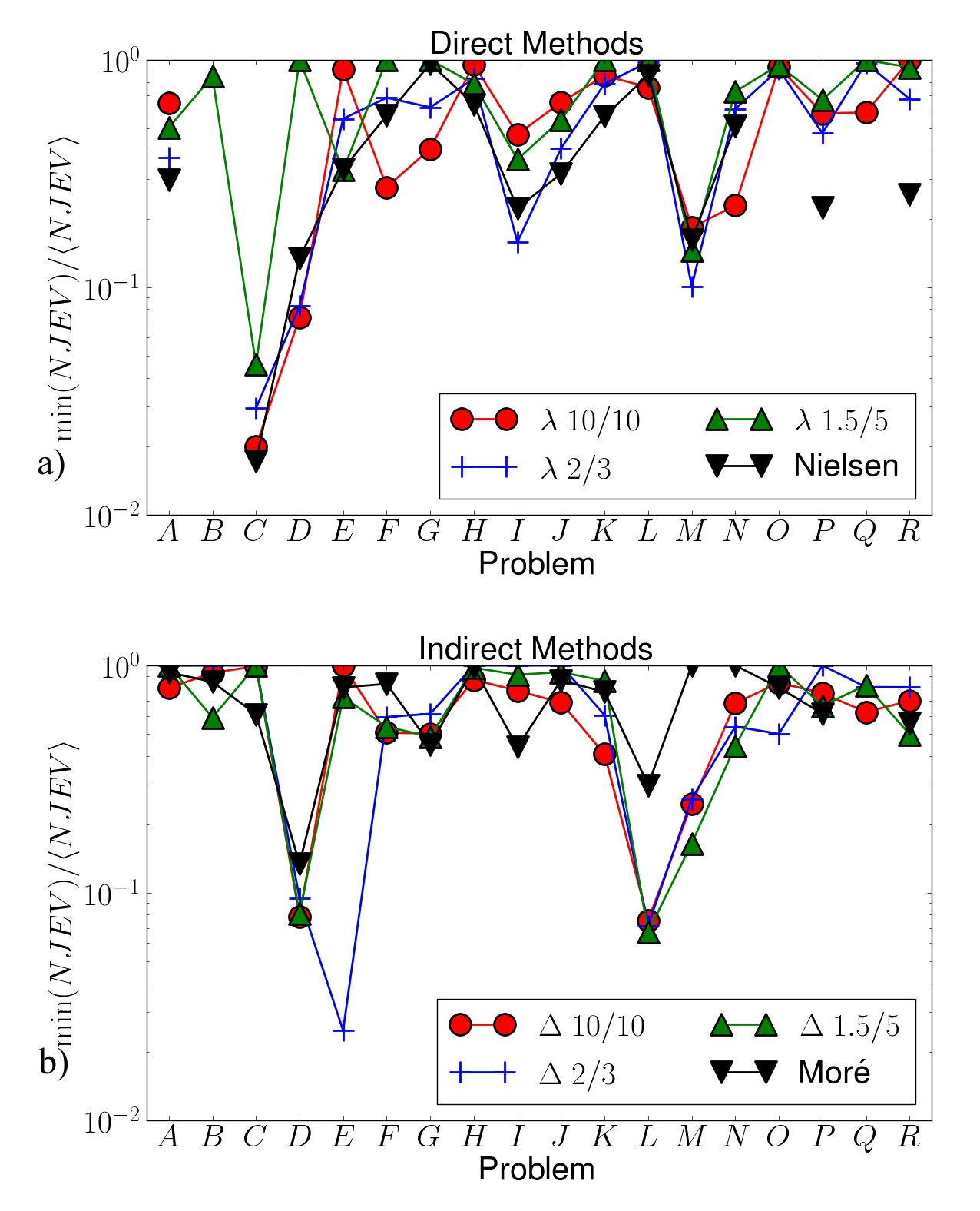}
  \caption[Relative inverse NJEV]{\textbf{Relative Inverse NJEV} of direct (a) and indirect (b) methods of choosing $\lambda$.  The relative inverse NJEV is found by dividing an algorithm's average inverse NJEV by the fewest average inverse NJEV of any algorithm.  Since we plot the inverse relative NJEV, larger numbers imply a more efficient algorithm.  Notice how the indirect methods are sluggish on problems D and L compared to the direct methods.}
  \label{LM:fig:njevrel}
\end{figure}

 The relative performance of the methods for selecting $\lambda$ described in section \ref{LM:sec:lambda}, together with more complicated schemes describe by Nielson \cite{nielsen1999} and Mor\'{e} \cite{More1977} are summarized in Figs.~\ref{LM:fig:successrel} - \ref{LM:fig:njevrel}.  In these figures we have labeled direct algorithms with the prefix $\lambda$ followed by the relative factors by which $\lambda$ is lowered and raised.  Similarly we have labeled indirect methods by the prefix $\Delta$ followed by the factor by which $\Delta$ is tuned.

Notice that no algorithm consistently outperforms the all the others on all the problems.  The relative success of any algorithm seems to depend very strongly on the problem.  However, there appear to be classes of problems for which indirect methods collectively outperform direct methods and vice versa.

\subsection{Comparison of methods with acceleration}
\label{LM:sec:accelcomparison}

\begin{figure}[htbp]
  \centering
  \includegraphics[height=7in]{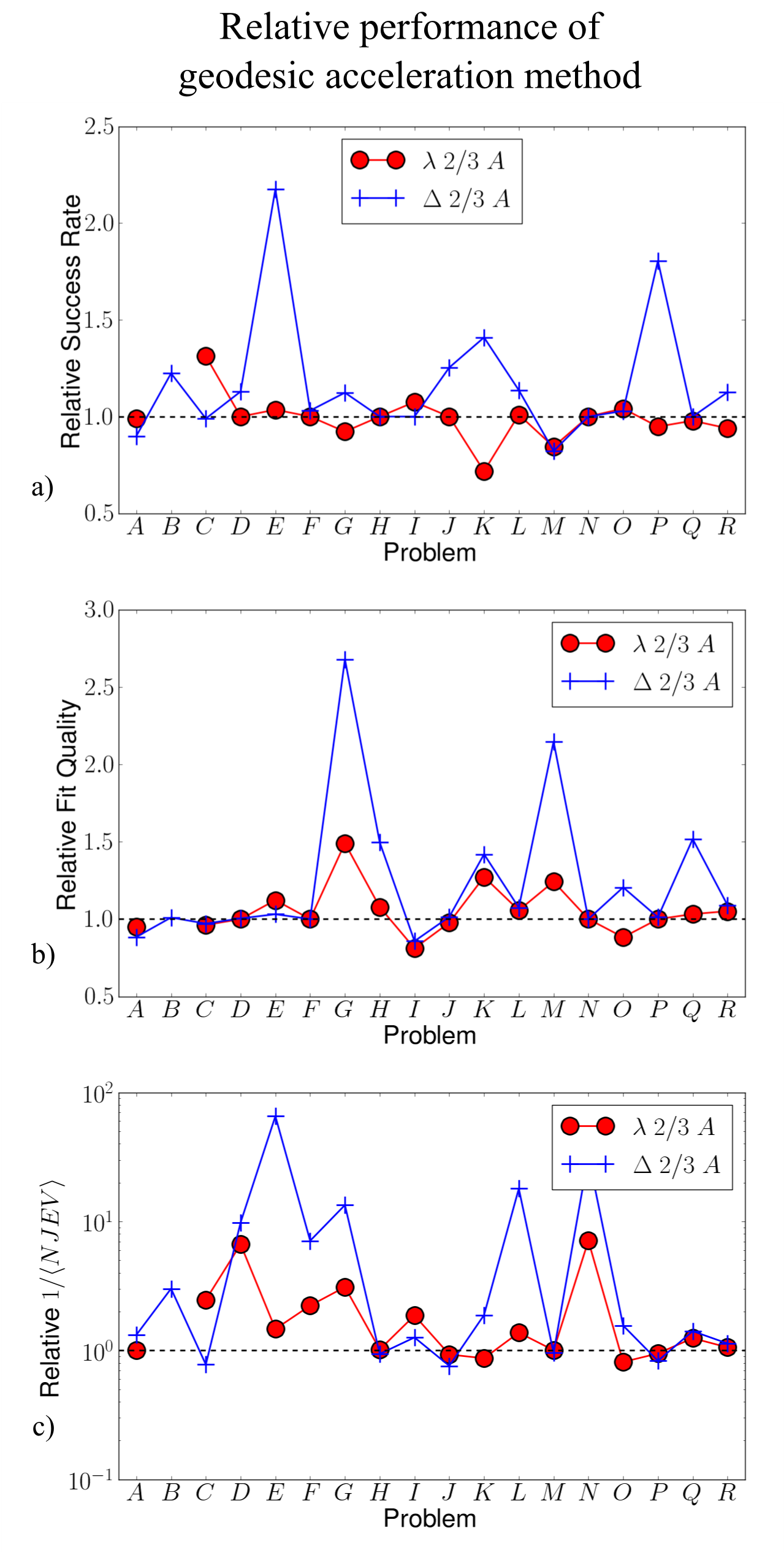}
  \caption[Performance of the geodesic acceleration algorithm]{\textbf{Performance of geodesic acceleration.}  The relative success rate (a), quality factor (b) and inverse NJEV (c) of two algorithms using geodesic acceleration.  The rates are each relative to each algorithm's performance without acceleration.  On each plot, points larger than $1$ (dashed black line) represent an improvement.  Notice that by including the acceleration the algorithm typically finds better fits more often in less time.  Perhaps most dramatically, in some cases the NJEV have been reduced by a factor of $70$!}
  \label{LM:fig:ResultsA}
\end{figure}

Figure \ref{LM:fig:ResultsA} summarizes the our test results for including geodesic acceleration.  Notice we have appended the letter $A$ after the algorithm to indicate the it now includes acceleration.  We see that, with a few exceptions, acceleration improves the algorithm's performance in each of the three measures we have used.  The most dramatic improvement is in the number of Jacobian evaluations necessary for convergence, where we saw speed ups as large as a factor of $70$, with most improvements between a factor of $2$ and $10$.

\subsection{Comparison of methods with bold acceptance}
\label{LM:sec:boldcomparison}

\begin{figure}[htbp]
  \centering
  \includegraphics[height=7in]{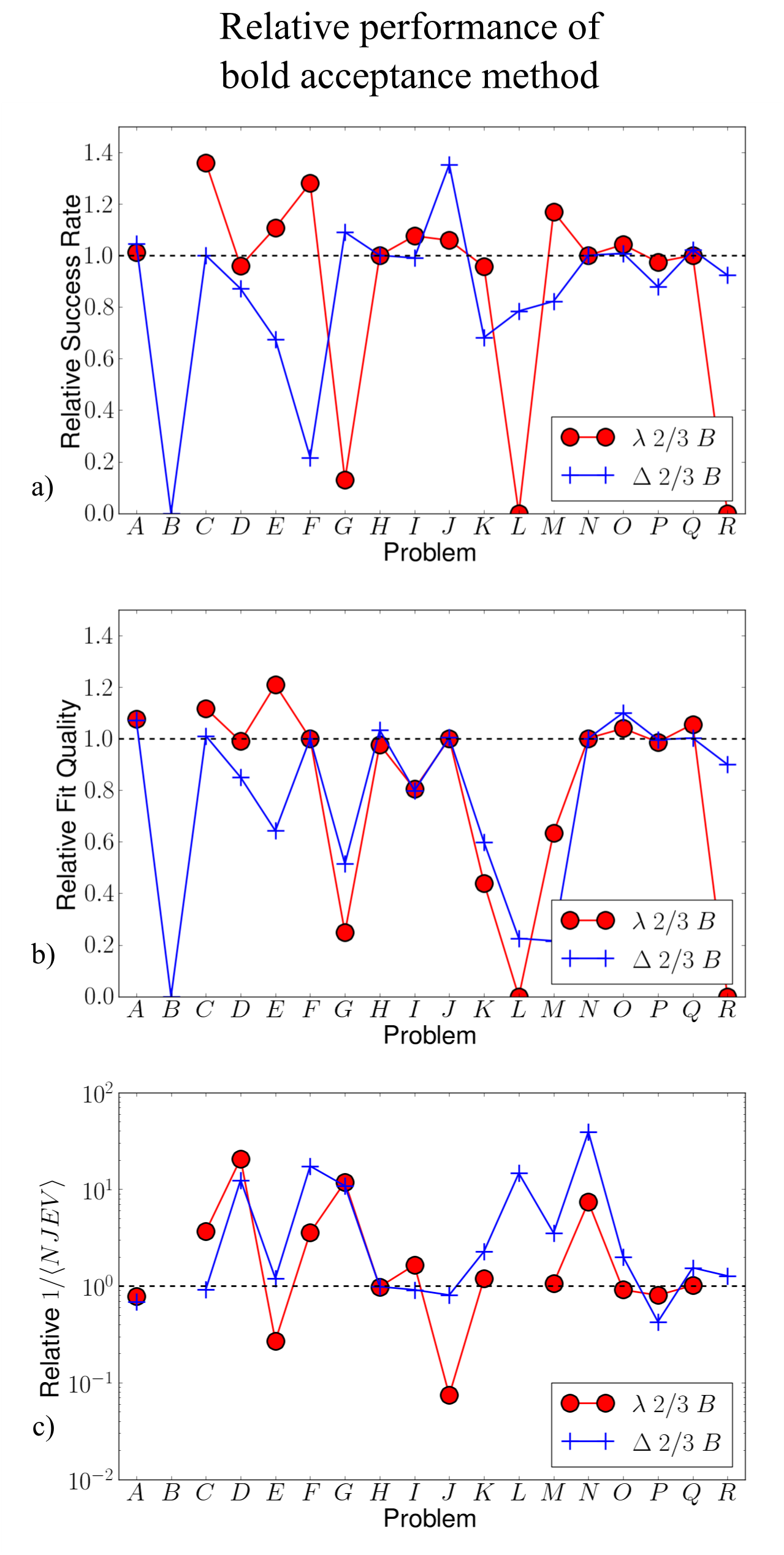}
  \caption[Performance of the bold method]{\textbf{Performance of bold acceptance}.  The relative success rate (a), fit quality (b) and inverse NJEV (c) of two algorithms using the bold acceptance criterion corresponding to Eq.~(\ref{eq.boldness2}) with $b=2$.  The rates are each relative to each algorithm's performance while accepting only downhill moves.  On each plot, points larger than $1$ (dashed black line) represent an improvement. For many of these problems, accepting uphill moves increases the probability that the algorithm will become lost.  However, when the algorithm does succeed, it may be much faster!}
  \label{LM:fig:ResultsB}
\end{figure}

\begin{figure}[htbp]
  \centering
  \includegraphics[height=7in]{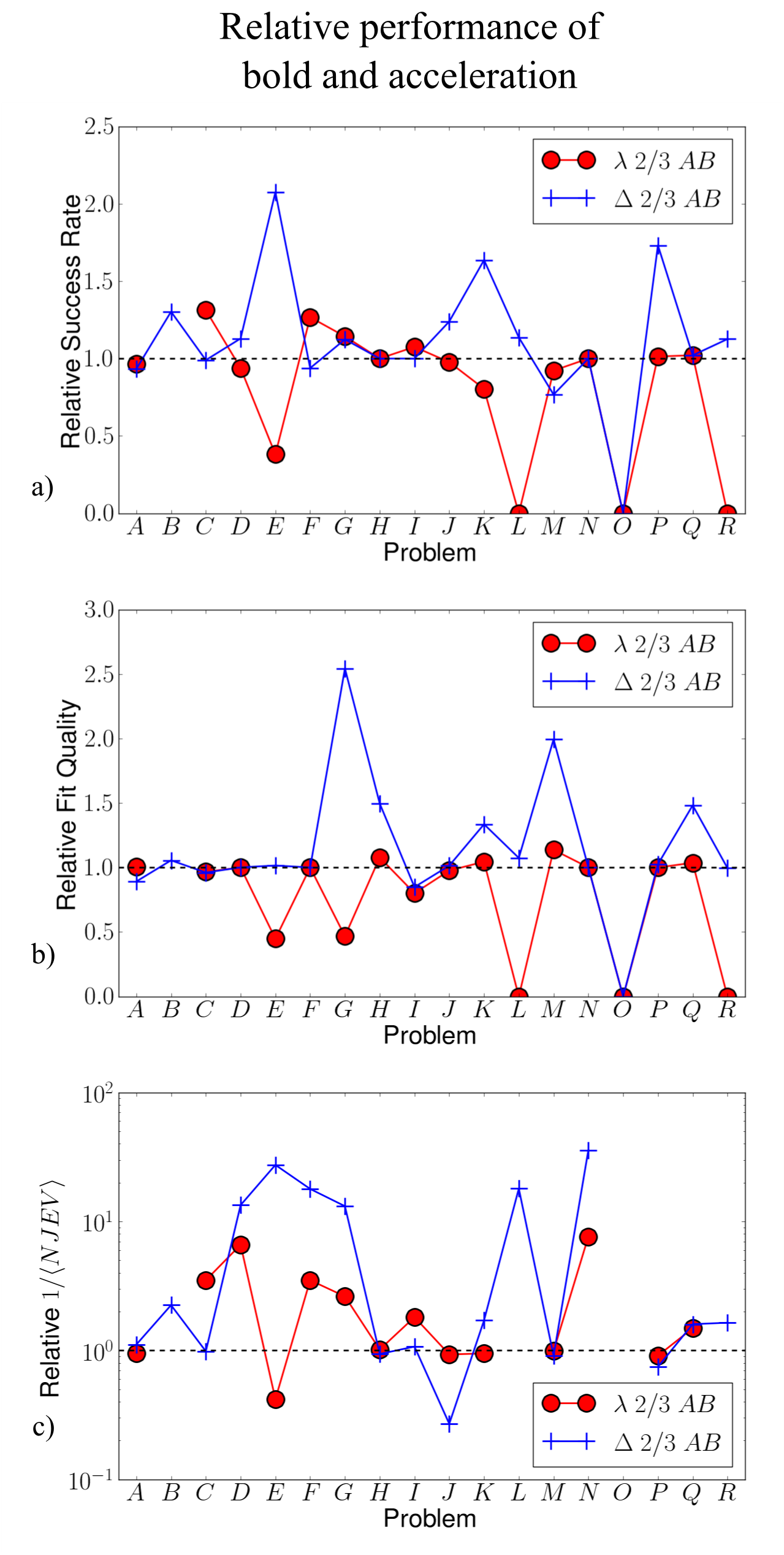}
  \caption[Performance of geodesic acceleration and bold]{\textbf{Performance with bold moves and acceleration.} The relative success rate (a), fit quality (b) and inverse NJEV (c) of two algorithms using bold acceptance criterion.  The rates are each relative to each algorithm's performance while accepting only downhill moves.  On each plot, points larger than $1$ (dashed black line) represent an improvement.  Notice that by including the acceleration we are able to prevent the algorithm from becoming lost when using the bold acceptance.}
  \label{LM:fig:ResultsAB}
\end{figure}

Figure \ref{LM:fig:ResultsB} summarizes our test results for including bold acceptance (using Eq.~(\ref{eq.boldness2}) with $b = 2$) to the standard methods.  Notice that we have appended the letter $B$ to the algorithm name to indicate that the bold method was used.  In figure~\ref{LM:fig:ResultsAB} we show results for combining geodesic acceleration with the bold acceptance method, which we denote by the letters $AB$.

\subsection{Comparison of methods with rank-deficient updates}
\label{LM:sec:boydencomparison}

\begin{figure}[htbp]
  \centering
  \includegraphics[height=7in]{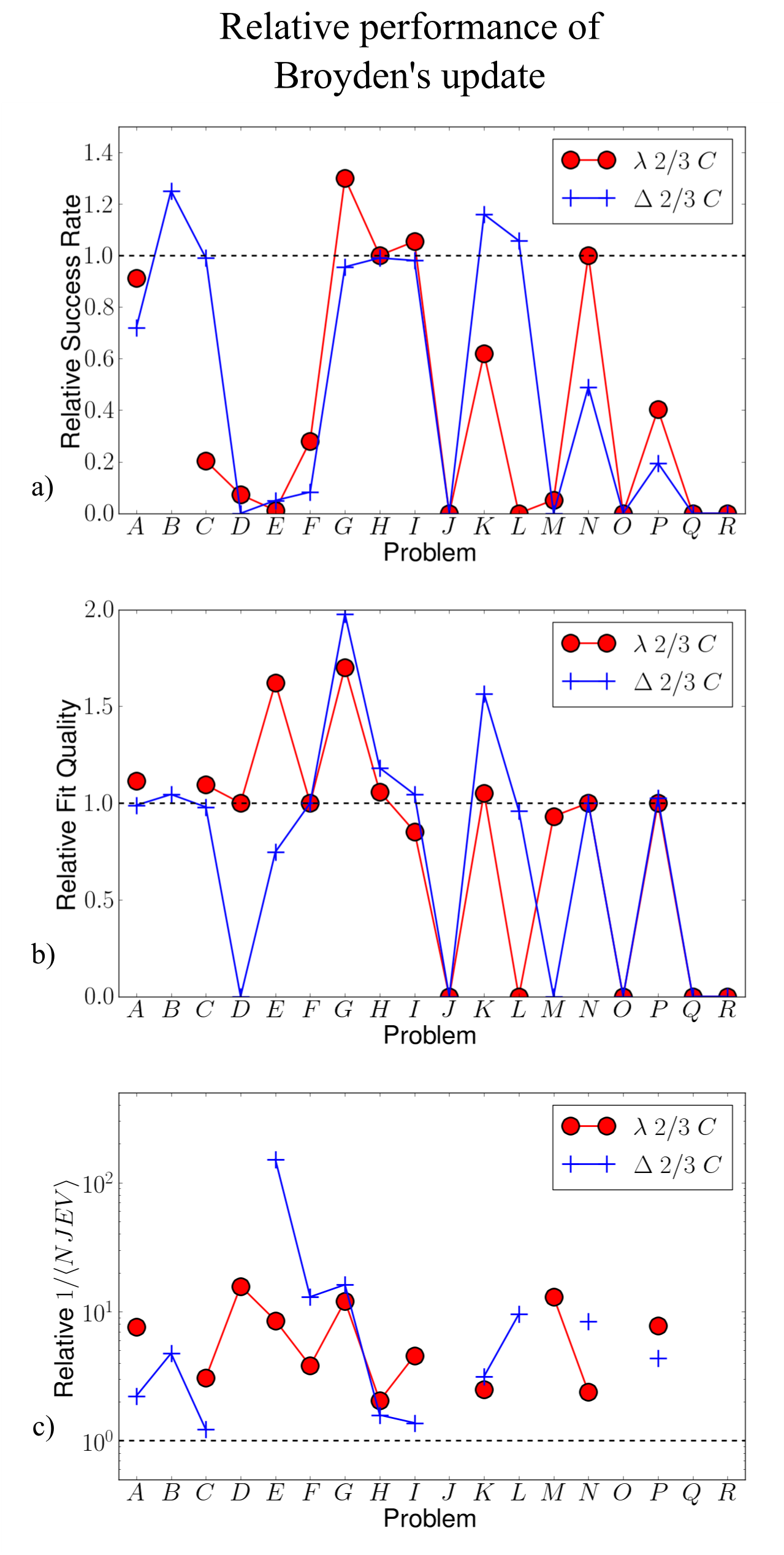}
  \caption[Performance using Broyden's update]{\textbf{Performance of Broyden's Update.}  The relative success rate (a), fit quality (b) and inverse NJEV (c) of two algorithms using a rank-$1$ Broyden's update.  The rates are each relative to each algorithm's performance without such an update.  On each plot, points larger than $1$ (dashed black line) represent an improvement.  Including the Broyden update typically causes the algorithm to be less robust to initial conditions, manifest by a lower success rate and average fit quality.  Without the need to reevaluate the Jacobian after each accepted step, the best fit can often be found much more quickly.  Note that missing points in (c) correspond to points for which the convergence rate and average quality factor was very near zero so that comparisons do not have any merit.}
  \label{LM:fig:ResultsC}
\end{figure}

\begin{figure}[htbp]
  \centering
  \includegraphics[height=7in]{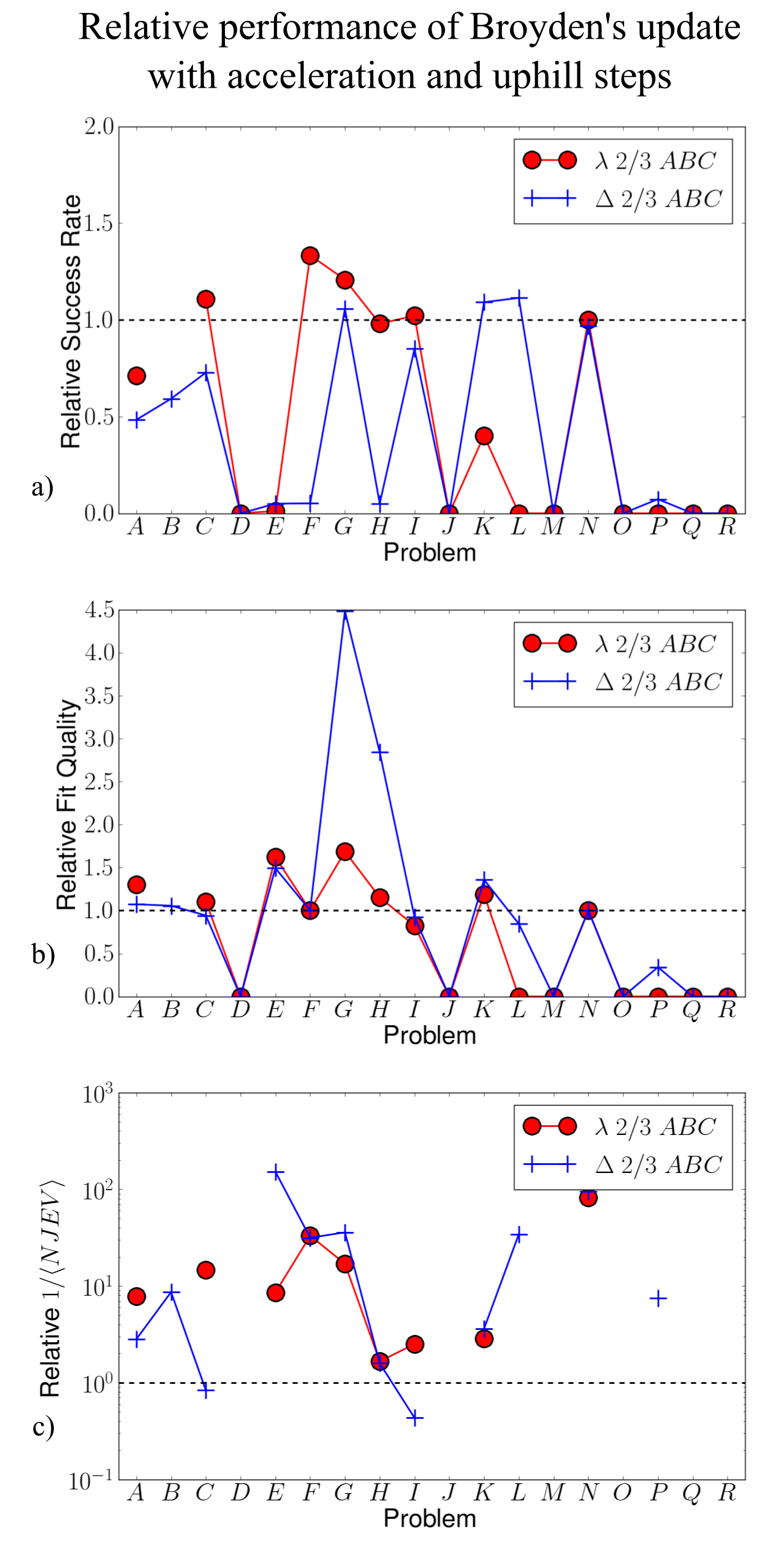}
  \caption[Performance using acceleration, bold, and Broyden's update]{\textbf{Performance of Broyden's Update with acceleration and bold moves.}  The relative success rate (a), fit quality (b) and inverse NJEV (c) of two algorithms using geodesic acceleration.  The rates are each relative to each algorithm's performance without acceleration.  On each plot, points larger than $1$ (dashed black line) represent an improvement.  Notice that by including acceleration and boldness does not have a strong effect on the results of the algorithm when using Broyden's udpate.}
  \label{LM:fig:ResultsABC}
\end{figure}

In figure~\ref{LM:fig:ResultsC} we present the performance results for methods with rank-$1$ updates after accepted steps. (Note that we distinguish a method using Broyden's update by the letter $C$ in the algorithm's name.)  In figure \ref{LM:fig:ResultsABC} we present the results of the algorithm that uses a rank-$2$ update using both geodesic acceleration and the bold acceptance method.

\bibliographystyle{model1a-num-names}
\bibliography{C:/Working/Z/References}







\end{document}